\documentclass[lettersize,journal]{IEEEtran}
\usepackage{amsmath,amsfonts}
\usepackage{algorithmic}
\usepackage{algorithm}
\usepackage{array}
\usepackage[caption=false,font=normalsize,labelfont=sf,textfont=sf]{subfig}
\usepackage{textcomp}
\usepackage{stfloats}
\usepackage{url}
\usepackage{verbatim}
\usepackage{graphicx}
\usepackage{multirow}
\usepackage{cite}
\usepackage{color}
\usepackage{hyperref}
\usepackage[mathscr]{eucal}

\DeclareRobustCommand\onedot{\futurelet\@let@token\@onedot}
\def\eg{\emph{e.g., }}  

\def\ie{\emph{i.e., }}

\def\etc{\emph{etc. }}

\def\etal{\emph{et al.} }
\makeatother

\begin{document}

\title{Segmentation Guided Sparse Transformer for Under-Display Camera Image Restoration}

\author{Jingyun Xue,
        Tao Wang,
        Pengwen Dai,
        Kaihao Zhang
\IEEEcompsocitemizethanks{\IEEEcompsocthanksitem J. Xue, W. Ren and X. Cao are with Shenzhen Campus of Sun Yat-sen University, Shenzhen, China, 
(e-mail: xuejy5@gmail.com, \{renwq3, caoxiaochun\}@mail.sysu.edu.cn).
\IEEEcompsocthanksitem W. Luo is with the Hong Kong University of Science and Technology, Hong Kong, China, and Shenzhen Campus of Sun Yat-sen University, Shenzhen, China, (e-mail: china@gmail.com).
\IEEEcompsocthanksitem T. Wang is with the State Key Laboratory for Novel Software Technology, Nanjing University, Nanjing, China, (e-mail: taowangzj@gmail.com).
\IEEEcompsocthanksitem J. Wang, Z. Liu are with Samsung Research China - Beijing (SRC-B), Beijing, China, (e-mail: \{jun01.wang, zikun.liu\}@samsung.com).
\IEEEcompsocthanksitem K. Zhang is with the Harbin Institute of Technology, Shenzhen, China, (e-mail: super.khzhang@gmail.com).
\IEEEcompsocthanksitem Hyunhee Park is with the Department of Camera Innovation Group, SAMSUNG Electronics, Suwon 16677, South Korea, (e-mail: inextg.park@samsung.com).
}
}

\markboth{Journal of \LaTeX\ Class Files,~Vol.~14, No.~8, August~2021}%
{Shell \MakeLowercase{\textit{et al.}}: A Sample Article Using IEEEtran.cls for IEEE Journals}

\maketitle
\begin{abstract}
Under-Display Camera (UDC) is an emerging technology that achieves full-screen display via hiding the camera under the display panel. 
However, the current implementation of UDC causes serious degradation. The incident light required for camera imaging undergoes attenuation and diffraction when passing through the display panel, leading to various artifacts in UDC imaging. 
Presently, the prevailing UDC image restoration methods predominantly utilize convolutional neural network architectures, whereas Transformer-based methods have exhibited superior performance in the majority of image restoration tasks. This is attributed to the Transformer's capability to sample global features for the local reconstruction of images, thereby achieving high-quality image restoration.
In this paper, we observe that when using the Vision Transformer for UDC degraded image restoration, the global attention samples a large amount of redundant information and noise.
Furthermore, compared to the ordinary Transformer employing dense attention, the Transformer utilizing sparse attention can alleviate the adverse impact of redundant information and noise. 
Building upon this discovery, we propose a Segmentation Guided Sparse Transformer method (SGSFormer) for the task of restoring high-quality images from UDC degraded images. Specifically,  we utilize sparse self-attention to filter out redundant information and noise, directing the model's attention to focus on the features more relevant to the degraded regions in need of reconstruction. Moreover, we integrate the instance segmentation map as prior information to guide the sparse self-attention in filtering and focusing on the correct regions.
Experimental results on public datasets verify that the proposed method demonstrates positive performance in comparison to state-of-the-art approaches.
\end{abstract}


\begin{figure}[t] 
\centering
\includegraphics[width=0.45\textwidth]{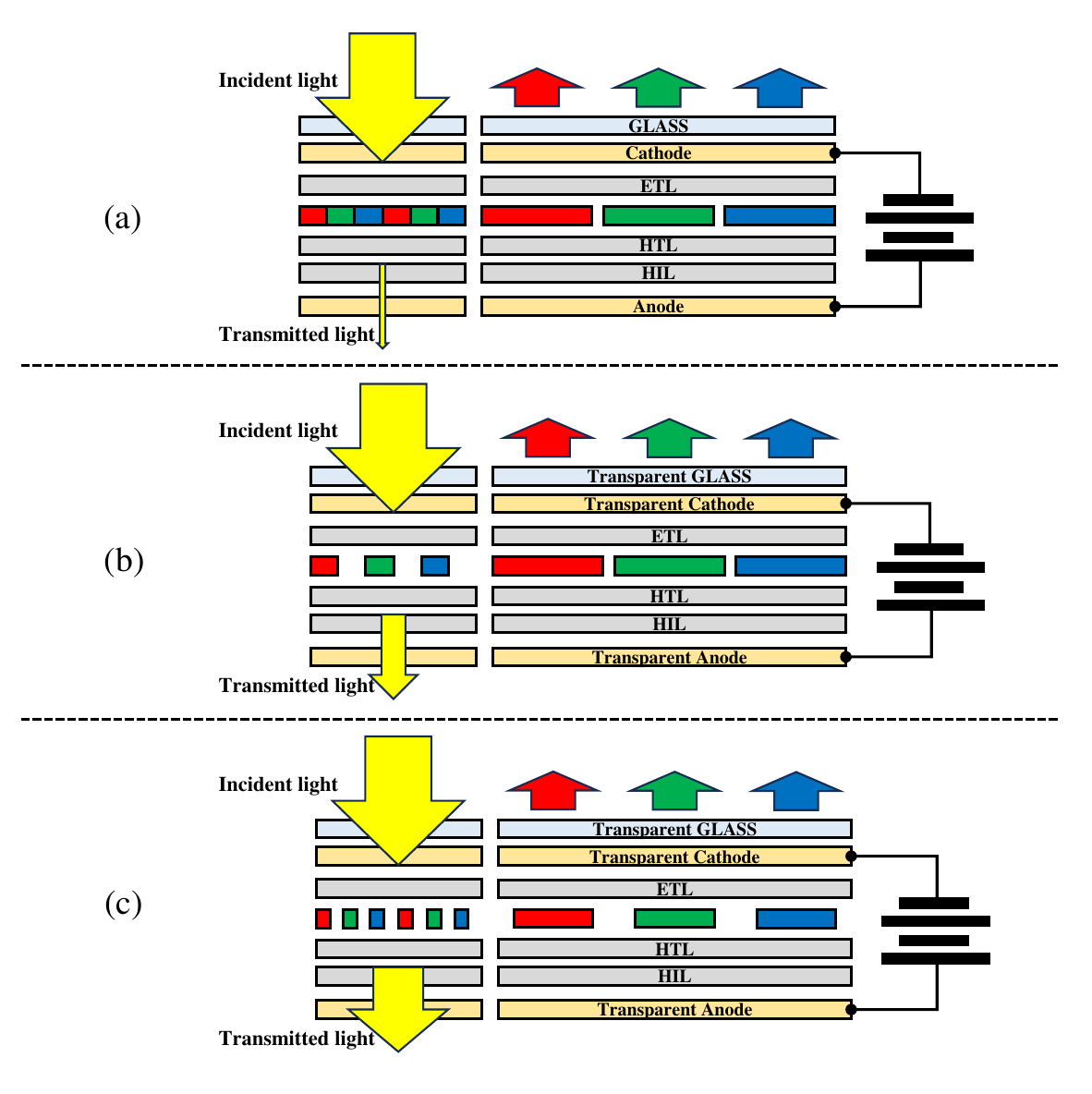}
\caption{The structural diagrams of regular OLED screens and two different forms of UDC screens, with the incident light direction indicated on the left, and the Light-Emitting direction on the right. 
(a) The fundamental structure of OLED screen panel. 
(b) The screen panel structure diagram of UDC is achieved by reducing the pixel density within the screen above the camera. (c) The screen panel structure diagram of UDC is achieved by using smaller pixel sizes within the screen above the camera.}
\label{fig1}
\end{figure}

\section{Introduction}
Nowadays, the prevalence of smart devices continues to rise, with most of these devices equipped with touchscreen displays and front-facing cameras, such as smartphones, tablets and the like. Serving as the direct interface between intelligent devices and users, the screen plays a pivotal role in determining the human-computer interaction experience of intelligent devices. Thus, for the purpose of superior display performance, device manufacturers have been consistently striving to improve the display area ratio of the screen. Under-display camera (UDC) is a novel front-facing camera form that maximizes the screen proportion via hiding the camera below the display panel, to implement full-screen display without punch holes or notches. UDC represents an effective approach for harmoniously integrating the screen and front-facing camera on smart devices, thus it has the potential to become a prevalent front-facing camera design scheme in the future.

However, achieving full-screen display through UDC comes at a cost: there would be serious degradation in UDC imaging. The incident light must pass through the screen panel before reaching the photosensitive elements of the camera's image sensor, which causes inevitable and serious degradation, \eg blur, noise, color aberration and low light, \etc\cite{yanshe, UDCIR, DISCNet}. 
To enhance the imaging performance of UDC, the commonly employed approach is to enhance the incident light intensity by reducing pixel density above the camera and ease the diffraction phenomena by adjusting the layout of RGB sub-pixels~\cite{pdl, pdl2, pdl3}.  
Compared to the hardware solution, adopting an image restoration neural network to repair UDC raw images is not only a more cost-effective but also a more remarkable approach compared to above approaches, and it is becoming the mainstream trend~\cite{csvt1,UDCIR, DISCNet,  kwon2021controllable, DAGF, mipi, udcresnet, PDCRN, ECCVchallenge, UDCUNet, qi2021isp, Nonaligned, wacv1,tjf}.


To investigate the UDC image restoration network, current research initiates by delving into the imaging principles of UDC to model the degradation phenomenon~\cite{UDCIR, DISCNet}. The fundamental structure and emitting pathway of OLED is illustrated in Fig.~\ref{fig1}~(a), showing that incident light undergoes substantial attenuation after passing through the regular nontransparent OLED screen panel, with a transmittance of less than 3\%~\cite{UDCIR}. There are currently two main methods for the industry to implement UDC~\cite{mix4}, as illustrated in Fig.~\ref{fig1}~(b) and Fig.~\ref{fig1}~(c). These two UDC implementation methods have consistent principles for image degradation~\cite{UDCIR, DISCNet}: color deviation and low light caused by light passing through multiple thin-film layers of the OLED panel, as well as diffraction artifacts caused by passing through the apertures between pixels.


Based on the imaging principles of UDC above, Zhou~\etal\cite{UDCIR} and Feng~\etal\cite{DISCNet} establish the physical model of the degradation of UDC images respectively, and then generate and release the datasets with paired high-quality images and corresponding UDC degraded images. Numerous approaches ~\cite{csvt1, UDCIR, DISCNet, kwon2021controllable, DAGF, mipi, udcresnet, PDCRN, ECCVchallenge, UDCUNet, qi2021isp, Nonaligned, wacv1, tjf} have been proposed for UDC image restoration tasks on the datasets mentioned above. However, the majority of methods are transferred from methods designed for other image restoration tasks, which generally do not fully consider the characteristics of UDC degraded images.
First, compared to single-type-degradation tasks such as denoising~\cite{restormer, uformer, chen2021pre, tu2022maxim, guo2015efficient}, deblurring~\cite{restormer, uformer, chen2021pre, wt1, tu2022maxim, wen2020simple}, low light image enhancement~\cite{wt2, tu2022maxim, li2021low}, deraining~\cite{restormer, uformer, chen2021pre, jiang2020multi, tu2022maxim, zhang2022beyond, zhang2019image, jiang2020decomposition}, and dehazing~\cite{tu2022maxim, qiu2023mb, zhang2020multi}, the degradation of UDC images is more intricate, including color deviation, blurriness, low light and glare, among other degradations~\cite{yanshe, UDCIR, DISCNet}. Thus, simple models or fixed invariant PSF cannot adequately capture the degradation in UDC images. 
Second, the restoration of UDC degraded images requires both the local texture information for reconstructing the high-intensity diffraction generated glare regions, as well as global offset characteristics for eliminating the systematic color deviation induced by panel thin-film filtering~\cite{UDCUNet}. 
Third, The application scenarios of UDC images determine that they are typically high-resolution images with limited feature information carried by individual pixels. Traditional global feature capture methods often capture a large amount of redundant information, resulting in resource wastage and decreased accuracy of global features.
Considering these characteristics of UDC image restoration tasks and the corresponding advantages of sparse networks, including direct attention towards key feature domains, reducing interference from redundant information and less computational overheads \cite{sparse2}, sparse networks represent a more promising solution.

\begin{figure}[t] 
\centering
\includegraphics[width=0.45\textwidth]{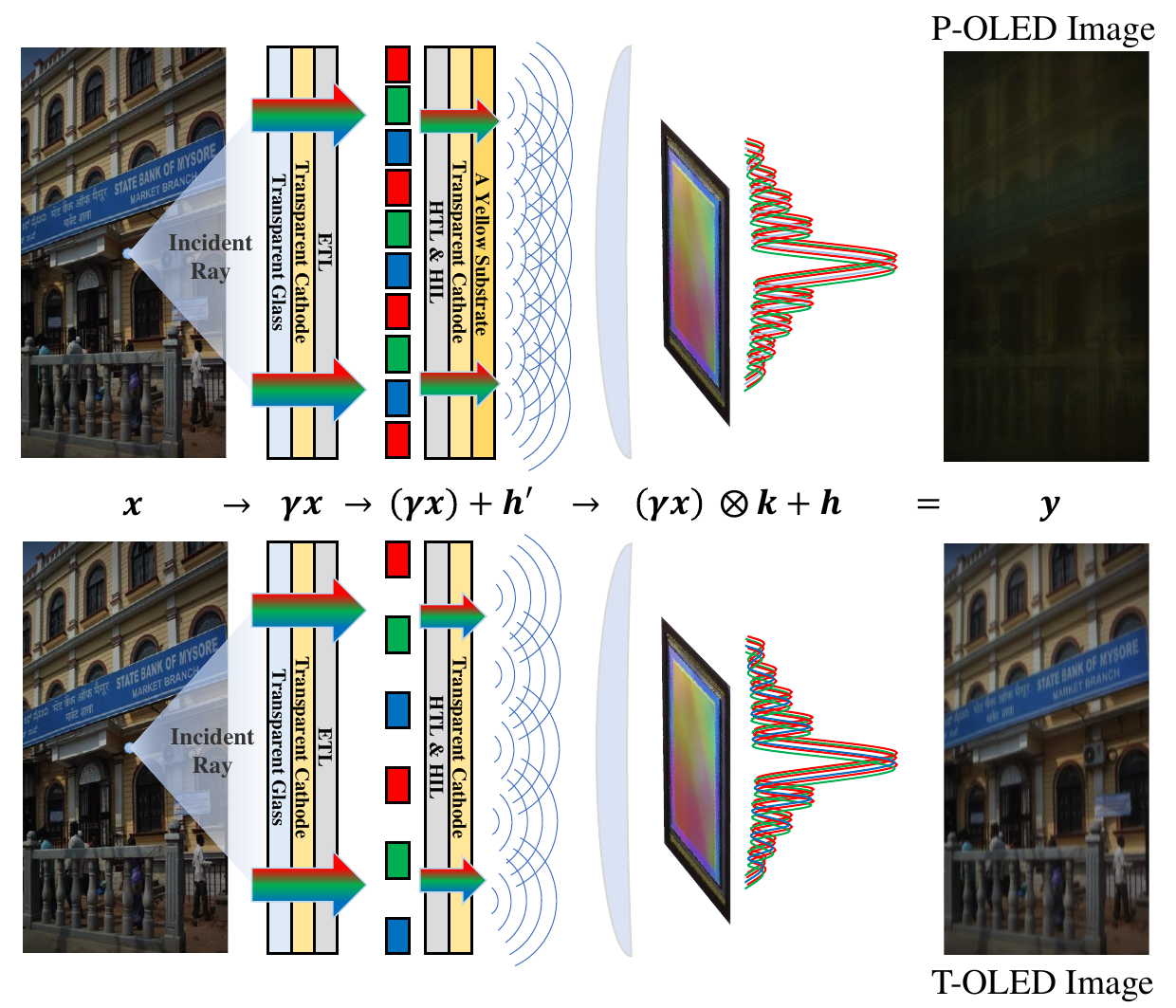}
\caption{The illustration of the UDC imaging system for P-OLED dataset and T-OLED dataset. In the optical system of P-OLED, sub-pixels are densely arranged on a yellow substrate, leading to low transmittance. Consequently, the imaging exhibits significant color deviation and low light degradation. In the optical system of T-OLED, the sub-pixels are sparsely arranged and there is no substrate, resulting in a high transmittance. The imaging is blurred caused by diffraction.}
\label{fig2}
\end{figure}


In this paper, we propose SGSFormer, a sparse Transformer-based algorithm for restoration of UDC images. The algorithm utilizes an instance segmentation map as guiding information to enhance the restoration process. 
First, building upon the aforementioned characteristics of degraded UDC images, we introduce a Transformer network to capture global color offset features and static blur offset features. Notably, the acquisition of static blur offset features can serve as a replacement for the domain knowledge provided by the Point Spread Function (PSF), facilitating blind PSF restoration. 
Furthermore, due to the abundant global redundant information in UDC degraded images, common vision Transformer networks often perform extensive redundant computations, compromising performance. To address this issue, we introduce sparse attention with the capability to filter redundant information and noise, as well as focusing attention on important features. Based on it we devise a sparse Transformer structure consisting of alternating dense attention and sparse attention to simultaneously consider both global and local features. 
Concurrently, we recognize that the performance of sparse attention can be enhanced by incorporating additional prior-guided information. Therefore, we propose Segmentation Guided Sparse Attention, which leverages instance segmentation features as guiding information to improve the performance of sparse self-attention. Specifically, the instance information refined by segmentation features guides the ranking strategy to improve filtering precision and enhance the accuracy of the Value Matrix through adjustments (Sec. \ref{sec:decoder}). 
Lastly, we design a lightweight version of Segmentation Guided Sparse Attention to streamline the network and implement an asymmetric design for the overall U-net framework. (Sec. \ref{sec:overall}).

The main contributions of this work can be summarized as follows:
\begin{itemize}
    \item We propose Segmentation Guided Sparse Attention, which utilizes instance segmentation features as guiding information to enhance the performance of sparse self-attention.
    \item We further propose SGSFormer, an asymmetric U-net network architecture that accommodates various processes within the network by employing Transformer blocks with different structures.
    \item Experimental results indicate that our SGSFormer demonstrates favorable performance in comparison to existing works.
\end{itemize}

\begin{figure*}[t]
\centering
\includegraphics[width=0.95\textwidth]{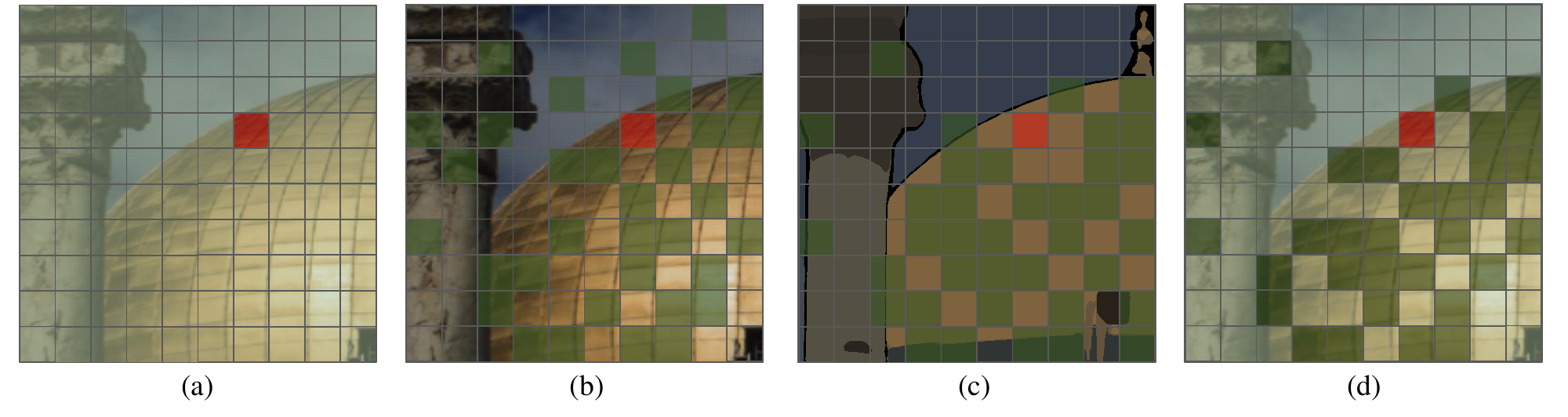}
\caption{Illustration of the various attention receptive fields. (a) Dense attention uniformly perceives each region of the global context. (b) Sparse attention selectively perceives regions with higher relevance. (c) Segmentation Guided Sparse Attention can better judge which regions have the higher correlation and then perceive the regions with the highest correlation. (d) Decoders employ Segmentation Guided Sparse Attention and dense attention alternately, thereby possessing a more comprehensive receptive field.}
\label{fig6}
\end{figure*}

\section{Related Work}
\label{sec:related_work}
\subsection{UDC Image Degeneration}
The physical modeling of UDC image degradation processes~\cite{pdl, pdl2, pdl3, yanshe} is primarily applied to analyze the diffraction pathways for the optimization of physical optical structures. Zhou~\etal\cite{UDCIR} are the first to introduce neural networks into the field of UDC image restoration, and also the first to analyze and model the degradation of UDC images from the perspective applicable to neural networks. Specifically, they replace the complex Fourier formula used for diffraction analysis with a forward model that can simulate PSF by using convolutional processes, so that they can approximate the blur kernel (\ie PSF) and consequently generate the UDC degeneration images datasets from ground-truth images. The model for the forward convolution process is expressed in the following formulas:
\begin{equation}
    \boldsymbol{Y}=(\gamma \boldsymbol{X})\otimes k + n,
\label{eq1}
\end{equation}
where $\boldsymbol{X}$ is the clean real-world image, and $\boldsymbol{Y}$ is the degraded UDC image. $\gamma$ is the intensity scaling factor, measured using linear regression to calculate the changing ratio of the average pixel values in simulating UDC operating mode. $k$ is the Point Spread Function (PSF) with multiple wavelengths. $n$ is the zero-mean signal-dependent noise and $\otimes$ represents the convolution operator.

Feng~\etal\cite{DISCNet} propose a new real-world UDC image formation model. This degradation model is given by
\begin{equation}
    \boldsymbol{\hat{Y}}=\emptyset[C(\boldsymbol{X}*k+n)],
\label{eq2}
\end{equation}
where $\boldsymbol{X}$ is the clean real-world image that has a high dynamic range (HDR), and $\hat{\boldsymbol{Y}}$ is the degraded UDC image. $k$ denotes the PSF. $n$ models the camera noise. $*$ is the 2D convolution operator. $C(\cdot)$ is the clipping operation, and $\emptyset(\cdot)$ represents the non-linear tone mapping function. Some works~\cite{Nonaligned, UDCUNet, kwon2021controllable, yang2023designing} also model the imaging principle of UDC differently to design their own multi-module UDC image restoration network, but they have the same formal structure and do not deviate from the scope of the mentioned two. They can be viewed as complementary to the mentioned two. 

Zhou~\etal\cite{UDCIR} propose the Monitor-Camera Imaging System (MCIS) to generate datasets and release two UDC image datasets, the Pentile-OLED (P-OLED) dataset and the 4K Transparent-OLED (T-OLED) dataset. The P-OLED dataset simulates the actual operating mode of UDC by setting the camera behind the Pentile-OLED screen panel and generates degraded images of UDC. The P-OLED panel has regular RGB-pixels and a yellow substrate, and the diagram of the imaging system is shown in Fig.~\ref{fig2}. The T-OLED dataset uses the same method of setting the camera behind the 4K Transparent-OLED transparent screen panel~\cite{transparent} to generate degraded images of UDC.
Its pixel distribution, as illustrated in Fig.~\ref{fig2}, is not dense but rather sparse.
Feng~\etal\cite{DISCNet} propose and release the synthetic dataset generated by the imaging formation model using the real-captured point spread function (PSF) of ZTE Axon20 phone's UDC images, which adopts the method of reducing the pixel density within the screen above the camera. Similar to the T-OLED operating mode, ZTE Axon20 phone's UDC uses a transparent cathode, transparent anode, and a transparent emitting layer.

\subsection{UDC Image Restoration}
Based on the datasets proposed by Zhou~\etal\cite{UDCIR} and Feng~\etal\cite{DISCNet}, a large amount of research commences to investigate the restoration of UDC degraded images using neural networks~\cite{kwon2021controllable, DAGF, mipi, udcresnet, PDCRN, ECCVchallenge, UDCUNet, qi2021isp, Nonaligned, wacv1, tjf}. In the ECCV 2020 challenge, some networks designed for other image restoration tasks have been adapted for the P-OLED and T-OLED datasets, and achieve excellent results~\cite{ECCVchallenge}. Also in the MIPI 2022 challenge, the neural networks transferred from other image restoration fields became mainstream and obtained the favourite performance~\cite{mipi}. Zhou~\etal\cite{UDCIR} use U-net~\cite{unet} and Resnet~\cite{resnet} to compare with the traditional general-purpose conventional pipeline (\ie Wiener Filter), and draw the conclusion that the neural network is superior. Inspired by the former, Yang~\etal\cite{udcresnet} meticulously design the Residual U-net and Dense U-net, achieving better results. Feng~\etal\cite{DISCNet} propose the DISCNet with multi-scale dynamic convolutions structure, utilizing generated filters that based on the PSF kernel codes. In the subsequent research, Feng~\etal\cite{Nonaligned} address the significant domain discrepancy and spatial misalignment issues by introducing an innovative Transformer-based framework, which produces accurately aligned and high-quality target data corresponding to the UDC input. Qi~\etal\cite{qi2021isp} propose an image-restoration pipeline independent of the Image Signal Processor (ISP), using a deep learning method that performs a RAW-to-RAW image restoration. Kwon~\etal \cite{kwon2021controllable} use pixel-wise UDC-specific kernel representation and a noise estimator to construct a controllable image restoration algorithm, resulting in higher perception. Liu~\etal\cite{UDCUNet} add the 2D dynamic information of the PSF into the U-net and design a tone mapping loss to achieve better performance. Koh~\etal\cite{BNUDC} propose a dual-branch network along with a preprocessing approach for the P-OLED dataset. And Song~\etal\cite{song2023under} also propose a two-branch restoration network. Conde~\etal\cite{wacv1} and Sundar~\etal\cite{DAGF} focus more on the smaller number of parameters for UDC image restoration networks, aiming for better deploy ability on mobile devices. Liu~\etal\cite{liu2023fsi} find that the degraded feature in the frequency domains is important and propose a frequency and spatial interactive learning network. Li~\emph{et al.}~\cite{csvt1} focus on model lightweight work. Tan~\emph{et al.}~\cite{tjf} use prior information for UDC image restoration.

\begin{figure*}[t]
\centering
\includegraphics[width=0.95\textwidth]{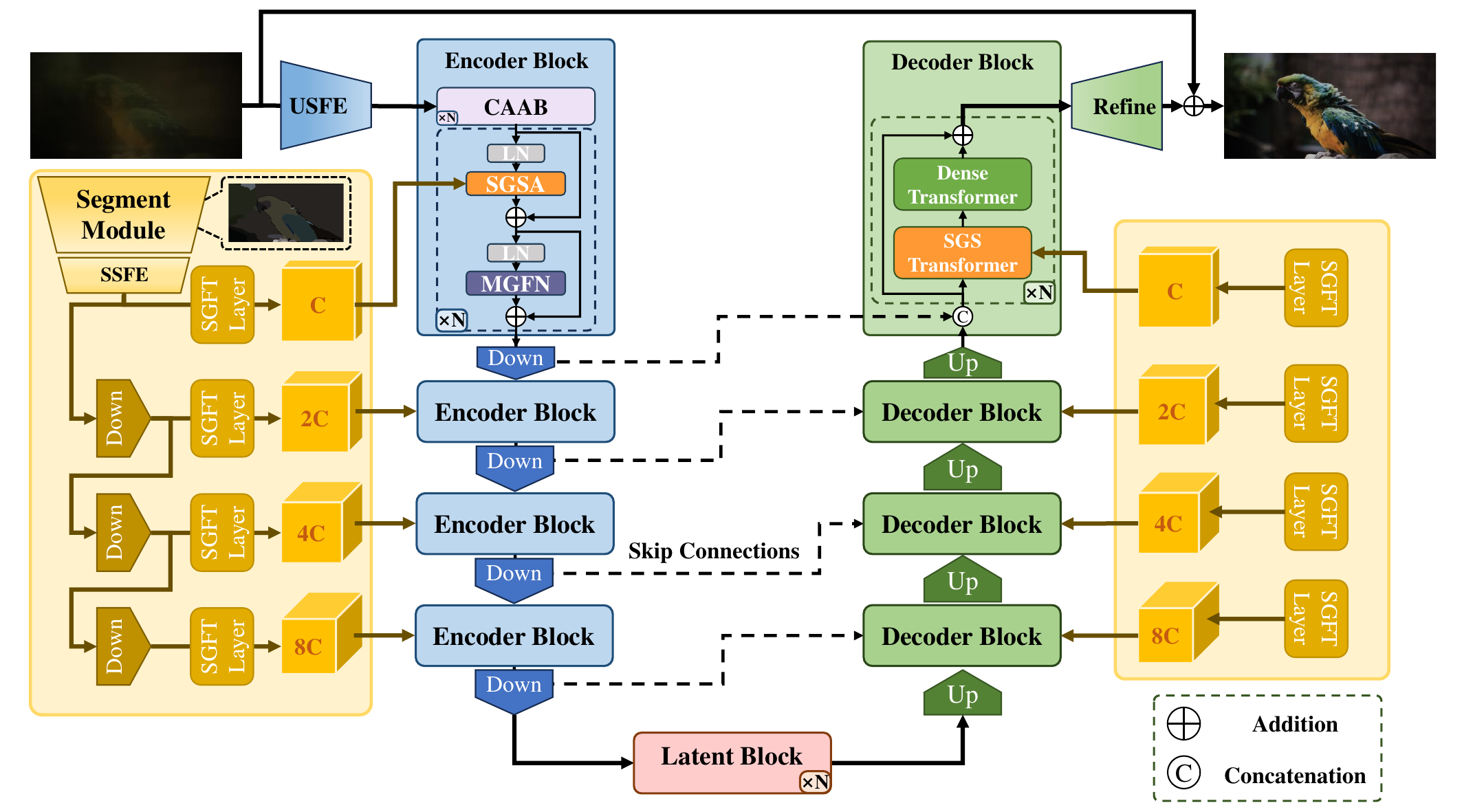}
\caption{Overview of the SGSFormer, which is an encoder-decoder architecture with residual connections. There are four encoder layers with down-sampling, four decoder layers with up-sampling, and a layer of latent block, represented by blue, green, and red squares respectively. The yellow square represents the Segmentation Feature Extraction Module, which provides different scales of segmentation features to each encoder and decoder.}
\label{fig4}
\end{figure*}

\subsection{Vision Transformer}
Driven by Transformer's significant success in the NLP domain~\cite{transformer}, Vision Transformer (ViT) has been rapidly advancing in recent years. After Dosovitskiy~\etal\cite{vit} first propose ViT by projecting large image patches into token sequences, ViT begins to be widely used in high-level vision tasks, such as image recognition~\cite{vit, recognition1, recognition2}, object detection~\cite{swint, carion2020end, twins} and segmentation~\cite{twins, zheng2021rethinking, wang2021pyramid, xie2021segformer}. The self-attention block in Transformer captures complex relationships between image patch sequences and establishes weight connections during image processing to better learn long-range dependencies. This is crucial for understanding the global structure and contextual information of images, so ViT achieves superior performance in high-level tasks~\cite{survey1, survey2}. Low-level tasks also benefit from these characteristics, such as super-resolution~\cite{swinir, chen2021pre}, deblurring~\cite{restormer, uformer, chen2021pre, wt1}, denoising~\cite{restormer, uformer, chen2021pre}, dehazing~\cite{qiu2023mb} and deraining~\cite{restormer, uformer, chen2021pre, jiang2020multi, zhang2022beyond, jiang2020decomposition}.

\subsection{Sparse Transformer}
Transformer comes with a high computational cost because the computational complexity of self-attention can increase quadratically with the number of image patches~\cite{transformer, sparse1}. Therefore, reducing the computational complexity of Transformer is crucial in high-resolution low-level tasks~\cite{restormer}. The sparse Transformer has been proposed as an effective method to reduce computational complexity in NLP~\cite{sparse1, sparse2, sparse4, sparse5}. Child~\etal\cite{sparse6} first propose sparse Transformer in Vision Transformer by employing sparse factorization of the attention matrix in self-attention, and decrease the computational complexity from order $O(n^2)$ to $O(n\sqrt{n})$. 
Liu~\etal\cite{sparse7} do not use a fixed sparse attention matrix, but instead use a dynamic pruning operator to implement the sparse Transformer. 
Fan~\etal\cite{sparse10} provide a different usage of sparse Transformer, which is to use Single-stride Sparse Transformer (SST) instead of down-sampling operation in LiDAR-based 3D object detection task. 
ART~\cite{sparse11} achieves SOTA performance in super-resolution using a combination of sparse and dense self-attention. Chen~\etal\cite{sparse12} develop an effective learnable top-k selection operator to implement the equivalent sparse Transformer, and~\cite{sparse0} propose a novel sparse sampling attention. 

\begin{figure}[t]
\centering
\includegraphics[width=0.45\textwidth]{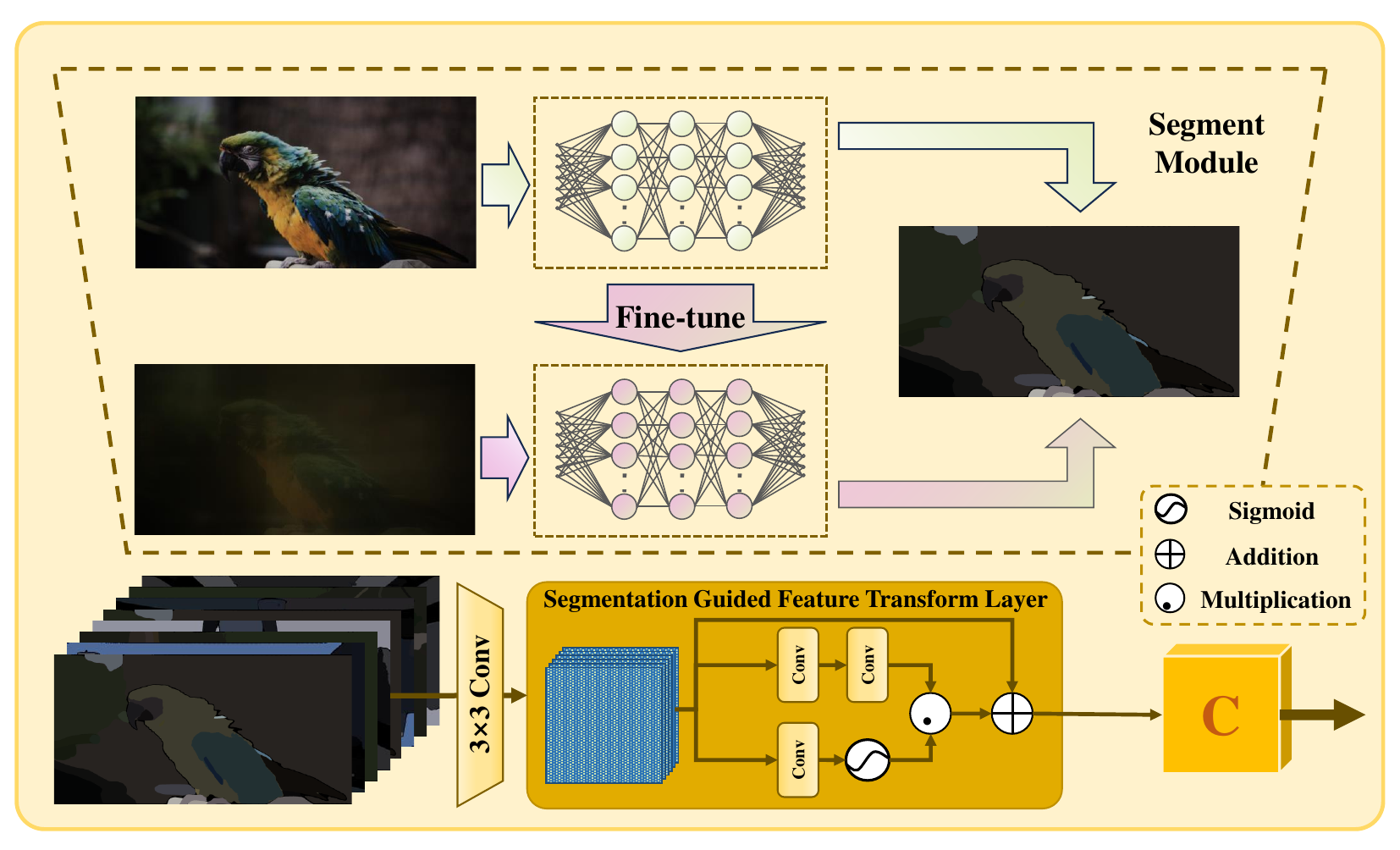}
\caption{The components of instance segmentation feature extraction module. The upper part is a fine-tuned instance segmentation network that outputs a segmentation map. The lower part is the feature transform layer, which transforms the segmentation map into multi-scale feature matrices and feeds them into the backbone network.}
\label{fig5}
\end{figure}

\section{Methodology}
\label{sec:methodology}

\subsection{Overall Pipeline}
\label{sec:overall}
The architecture of SGSFormer is illustrated in Fig.~\ref{fig4}, which is based on a hierarchical encoder-decoder framework with residual connection, \ie U-net structure. The start of the pipeline is to input a UDC degraded image $\boldsymbol{I_{udc}}\in\mathbb{R}^{H\times W\times 3}$, and then the instance segmentation networks generate the instance segmentation maps $\boldsymbol{I_{seg}}\in\mathbb{R}^{H\times W\times 3}$. Subsequently, maps $\boldsymbol{I_{seg}}$ are forwarded through the Instance Segmentation Feature Extraction Module and turn into the instance feature matrices, serving as the guiding information inputted into each Encoder and Decoder. The UDC degraded image $\boldsymbol{I_{udc}}\in\mathbb{R}^{H\times W\times 3}$, after passing through four sets of encoder blocks and down-sampling blocks, is input into the latent block, which consists of $N$ generic Transformer blocks. Following the latent block, the transformed features are passed through four sets of up-sampling blocks and encoder blocks. At the final stage of the pipeline, after the fusion of decoding features at different scales through the refine module, the repaired image is obtained via residual connections.

Unlike the symmetric structure of the traditional U-net, the proposed network employs an asymmetric structure to cater to the diverse requirements of different stages in the network. Specifically, encoder blocks are responsible for encoding multi-scale image features. Due to the limited amount of semantic information carried by individual pixel of degraded UDC images, it has a low-density requirement for global information sampling. Therefore, we adopt the Transformer composed entirely of the lightweight version of Segmentation Guided Sparse Attention. To refine the features of low-scale and high semantics, latent blocks should possess a comprehensive global receptive field. Hence, we deploy a common vision Transformer structure composed entirely of common attention. The decoder block is responsible for reconstructing images at different scales, with significant computational demands, necessitating the simultaneous consideration of global and local features. The sparse Transformer structure, which consists of alternating dense attention and Segmentation Guided Sparse Attention, is an appropriate choice.

\subsection{Instance Segmentation Feature Extraction Module}
\label{sec:segmentation}
Previous works~\cite{sg1, sg2, sg3} that adopt instance segmentation maps or semantic segmentation maps as prior information to enhance networks find the role of segmentation in guiding the convolutional kernels to effectively utilize the informative regions for restoring the degraded areas. Our work innovatively discover that instance segmentation maps also play a guiding role during the self-attention calculation of the sparse Transformer. In the proposed network, instance segmentation features act as prior knowledge to enhance the efficiency and accuracy of encoders in extracting and refining features at multiple scales, while also serving as constraint information to guide the sparse Transformer network in reconstructing the degraded portions of the UDC image in decoders.

\noindent\textbf{Segment Module.} 
In order to conveniently and expeditiously validate our ideas, we choose a network designed for zero-shot instance segmentation~\cite{sa}, named Segment Anything Model (SAM), rather than specifically training a domain-specific segmentation network. SAM generates a set of masks $\boldsymbol{\mathcal{M}_{udc}}$, with each mask corresponding to a segmentation object. We calculate the color mean values of the RGB channels for each segmentation object and assign them to the corresponding masks. Subsequently, we apply a weighted summation of all masks with color information onto the image $\boldsymbol{I_{udc}}$, resulting in $\boldsymbol{I_{seg}}$. The process is formulated as
\begin{equation*}
\boldsymbol{\mathcal{M}_{udc}} = Seg(\boldsymbol{I_{udc}}),
\end{equation*}
\begin{equation}
\boldsymbol{I_{seg}} = \alpha \boldsymbol{I_{udc}}+(1-\alpha)\sum_{\boldsymbol{M_i}\in\boldsymbol{\mathcal{M}_{udc}}}\mathcal{C}olor(\boldsymbol{M_i}).
\label{eq3}
\end{equation}
Based on the official release checkpoint, we conduct domain fine-tuning on small-scale UDC datasets to enhance the network's accuracy, transforming the instance segmentation task from zero-shot scenery to few-shot scenery. 
The datasets consist of input-output pairs, with the output being a set of masks generated by SAM on ground truth images, and the input being degraded UDC images, \ie $\{\boldsymbol{I_{udc}},\boldsymbol{\mathcal{M}_{gt}}\}$. 

Unlike SAM~\cite{sa} uses the weighted sum of Focal loss~\cite{lin2017focal} and Dice loss~\cite{diceloss}, we use the ``DiceCELoss'', which computes both the Dice loss and the cross-entropy loss and returns the weighted sum of these two losses. 

\noindent\textbf{Segmentation Guided Feature Transform Layer.} 
After obtaining the instance segmentation maps $\boldsymbol{I_{seg}}$ through the SAM fine-tuned network, this module will also extract feature matrices of multi-scale to directly guide the encoders and decoders. As shown in Fig.~\ref{fig5}, subsequent to the Segmentation Shallow Feature Extractor (SSFE), instance segmentation maps $\boldsymbol{I_{seg}}$ transform into the feature tensor $\boldsymbol{T_1}\in\mathbb{R}^{H\times W\times C}$. The down-sampling block is responsible for sampling feature tensor $\boldsymbol{T_1}$ to obtain the higher-level feature tensors $\boldsymbol{T_{2,3,4}}$ with higher semantics and less noise, while the Segmentation Guided Feature Transformer (SGFT) Layer is responsible for modulating the feature tensors $\boldsymbol{T_{1,2,3,4}}$ into modulating feature matrices $\boldsymbol{S_{1,2,3,4}}$ through the mapping function as
\begin{equation}
    \boldsymbol{T_1}=f_{1\times1}^{conv}(\boldsymbol{I_{seg}}),  \boldsymbol{T_i}=down(\boldsymbol{T_{i-1}})_{i\in\{2,3,4\}},
\label{eq7}
\end{equation}
\begin{equation}
    \boldsymbol{S_i}=\sigma(f_{1\times1}^{conv}(\boldsymbol{T_i}))\odot f_{1\times1}^{conv}(f_{1\times1}^{conv}(\boldsymbol{T_i})) \oplus \boldsymbol{T_i},
\label{eq8}
\end{equation}
where $\sigma$ is the sigmoid function. $f_{1\times1}^{conv}(\cdot)$ denotes a $1\times1$ convolution. $\odot$ is the element-wise multiplication operation and $\oplus$ is the addition operation.

\begin{figure*}[t]
\centering
\includegraphics[width=0.95\textwidth]{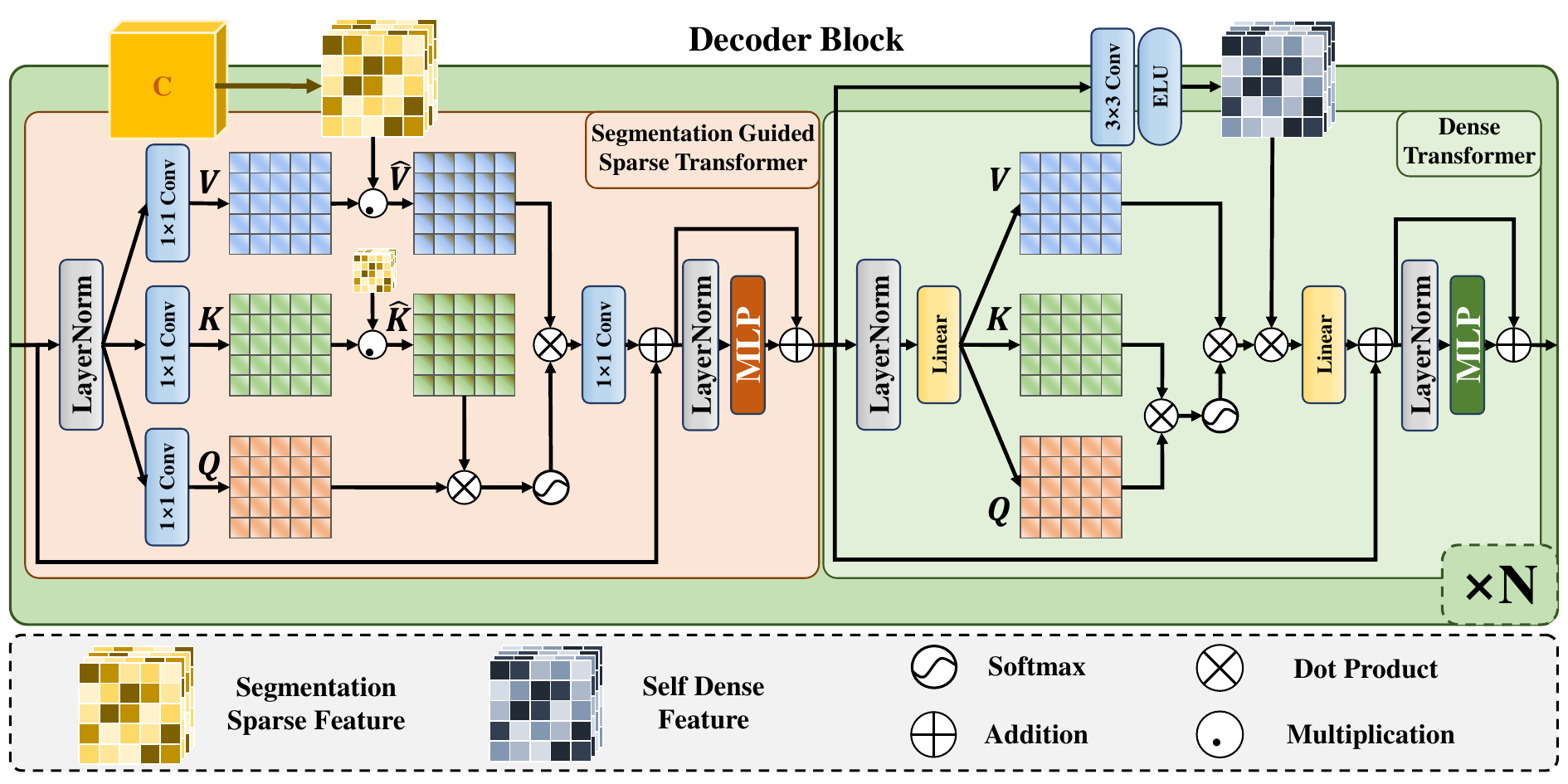}
\caption{Detailed structure diagram of the decoder. A decoding block is composed of N alternating Segmentation Guided Sparse Transformers and Dense Transformers. The primary distinctions between the two kinds of Transformers are as follows: (1) The former employs sparse attention, while the latter utilizes dense attention; (2) The former incorporates segmentation features into the Key matrix and Value matrix as guiding information, whereas the latter encodes itself as guiding information.}
\label{fig7}
\end{figure*}

\subsection{Decoder Block}
\label{sec:decoder}
Decoder blocks with multi-scale are responsible for reconstructing degraded regions of UDC images from high-level feature output by latent blocks. 
Previous studies~\cite{restormer,uformer} have achieved remarkable success by employing Vision Transformer, which possesses a global receptive field and strong contextual feature capturing capabilities, in decoder blocks for reconstructing the degraded regions using multi-scale features. 
However, while global self-attention effectively utilizes more comprehensive contextual features for image reconstruction, it also takes into consideration an excessive amount of noisy features and irrelevant redundant information. Consequently, impurity information is introduced during the reconstructing process, leading to a degradation in performance~\cite{sparse11,sparse12}. This deficiency is more pronounced in the task of UDC image restoration, as the feature information carried by individual pixels in UDC images is limited. Consequently, the vanilla Transformer's global receptive field introduces more redundant information. To address this issue, inspired by~\cite{sparse11, jiang2020decomposition}, we devise a Transformer paradigm consisting of alternating components: ``Segmentation Guided Sparse Transformer" and ``Dense Transformer", as illustrated in Fig.~\ref{fig4}. This paradigm can reduce the introduction of redundant information and noise without compromising the global receptive field. The following elucidates the constituent elements of this paradigm in detail.

\noindent\textbf{Segmentation Guided Sparse Transformer.} 
In order to mitigate the adverse effects of redundancy information and noise, we replace global self-attention with sparse self-attention. Inspired by~\cite{sparse12,sparse0}, this approach employs the top-k ranking strategy~\cite{topk} to filter out redundant information and noise, which facilitates the reconstruction process to concentrate on highly correlated features that are beneficial for reconstructing degraded regions. Furthermore, we use feature tensors transformed from instance segmentation maps to enhance the sparse self-attention block, named Segmentation Guided Sparse Attention (SGSA). Specifically, the instance information of relevance carried by the segmentation features guides the top-k ranking strategy to improve the filtering precision. Simultaneously, the relevant instance information also contributes to enhancing the accuracy of the Value Matrix through adjustments. The reconstruction Transformer block that employs SGSA is referred to as Segmentation Guided Sparse Transformer (SGS-Transformer).

SGSA differs from a common fully connected self-attention, which processes all Query-Key pairs indiscriminately with the Value matrix. Inspired by \cite{sparse12}, SGSA employs the top-k ranking strategy, selectively conducting calculations on Query-Key pairs and the Value matrix to achieve the purpose of filtering out redundant information and noise. Specifically, the top-k ranking strategy sorts the calculation results of each Query-Key pair, retaining those ranked higher than k and setting values ranked lower than k to zero. This makes that the rankings below k yield a result of zero in the subsequent computation with the Value matrix, thereby achieving the objective of discarding irrelevant information. We can consider it as a powerful activation function before $softmax(\cdot)$. The top-k ranking strategy $\mathcal{T}_k(\cdot)$ can be defined as:
\begin{equation}
    \mathcal{T}_k\left(\frac{\boldsymbol{QK}^T}{\sqrt{\lambda}}\right) = 
    \left\{\begin{aligned}
    &\mathcal{T}_k(\cdot) &rank(\mathcal{T}_k(\cdot)) > k_{th}\\
    &0 &otherwise\\
    \end{aligned}\right.
\label{eq9}
\end{equation}

Furthermore, SGSA uses instance segmentation features to guide the top-k ranking strategy to improve the rank precision. Unlike previous works~\cite{topk,topk2,topk3,sparse12} use simple weighted averages for sorting, whose accuracy has not been proven. The transformed instance segmentation features contain a wealth of relevant information, and their integration into the top-k ranking strategy can significantly enhance the scores and rankings of pertinent region features. Fig.~\ref{fig6} (b), (c) show the ranking results of sparse attention without segmentation guidance and SGSA, and we can find that SGSA ranks the regions with higher correlation in the correct position. Specifically, we perform element-wise multiplication between the instance segmentation features map and the Key matrix $\boldsymbol{K}$, resulting in the $\boldsymbol{\hat{K}}$. Similarly, we perform the same operation on Value matrix $\boldsymbol{V}$ to obtain $\boldsymbol{\hat{V}}$. $\boldsymbol{\hat{K}}$ and $\boldsymbol{\hat{V}}$ contain more relevant information, consequently exhibiting superior performance in the top-k ranking strategy as compared to $\boldsymbol{K}$ and $\boldsymbol{V}$. The above process is shown in the red section of Fig.~\ref{fig7}. The SGSA can be formulated as,
\begin{equation}
    \mathcal{SA}(\boldsymbol{D_i,S_i})=softmax\left[\mathcal{T}_k\left(\frac{\boldsymbol{Q}\left(\boldsymbol{K}\odot \boldsymbol{S_i}\right)^T}{\sqrt{\lambda}}\right)\right](\boldsymbol{V}\odot \boldsymbol{S_i}),
\label{eq10}
\end{equation}
where $\boldsymbol{D_i}$ represents the input, and Value matrix $\boldsymbol{V}$. $\boldsymbol{S_i}$ is the instance segmentation features map. $\odot$ denotes the element-wise multiplication and $\mathcal{T}_k(\cdot)$ denotes the top-k ranking strategy.

\begin{figure}[t]
\centering
\includegraphics[width=0.45\textwidth]{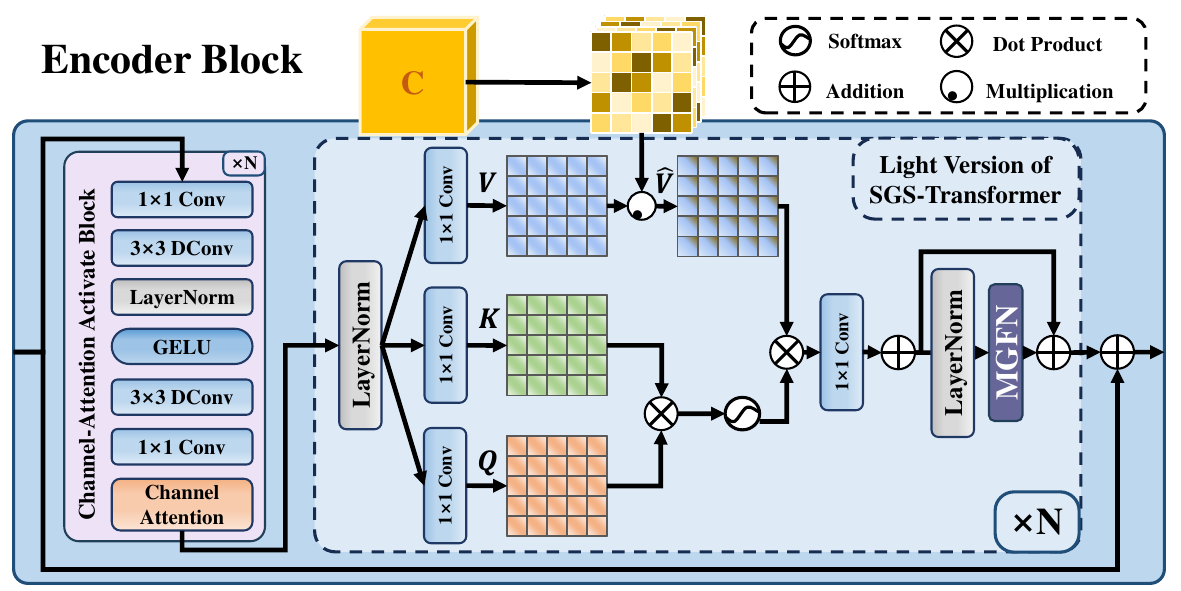}
\caption{Detailed structure diagram of the encoder. The purple section is the Channel Attention Activate Block, while the blue section represents the lightweight version of Segmentation Guided Sparse Transformer.}
\label{fig8}
\end{figure}

\noindent\textbf{Dense Transformer.} 
Nevertheless, relying solely on sparse attention during the reconstruction process may overcorrect, causing the attention map to overly focus on features sampled locally while neglecting the generic degenerative features within the global domain~\cite{sparse11}, such as global color offset features and static blur offset features. Inspired by~\cite{sparse11, sparse0}, we design a Dense Transformer with dense attention to focus on features within the global domain for image reconstruction. We alternate apply the aforementioned two Transformers to construct a reconstruction module, and reconstruct the degraded image under the combined action of the global receptive field and the local receptive field. 

The majority of the structure of the Dense Attention is the same as the traditional vision Transformer. Differently, to enhance attention's perception of global information, we introduce the self-dense feature in dense attention. This feature is derived from the input feature map and undergoes dot product with the final result. The above process is illustrated in the green section of Fig.~\ref{fig7}. The dense attention can be formulated as
\begin{equation*}
    \boldsymbol{S_i'} = \tau\left(f^{conv}_{3\times3}(\boldsymbol{D_i})\right),
\label{eq11.6}
\end{equation*}
\begin{equation}
    \mathcal{DA}(\boldsymbol{D_i,S_i'})=softmax\left(\frac{\boldsymbol{QK}^T}{\sqrt{\lambda}}\right)(\boldsymbol{V}\otimes \boldsymbol{S_i'}),
\label{eq11}
\end{equation}
where $\boldsymbol{D_i}$ represents the input, and $\boldsymbol{S_i'}$ is the self-feature map. $\otimes$ denotes the dot product and $\tau$ denotes ELU activation. $f^{conv}_{3\times3}$ is a $3\times3$ convolution.

\begin{figure}[t]
\centering
\includegraphics[width=0.45\textwidth]{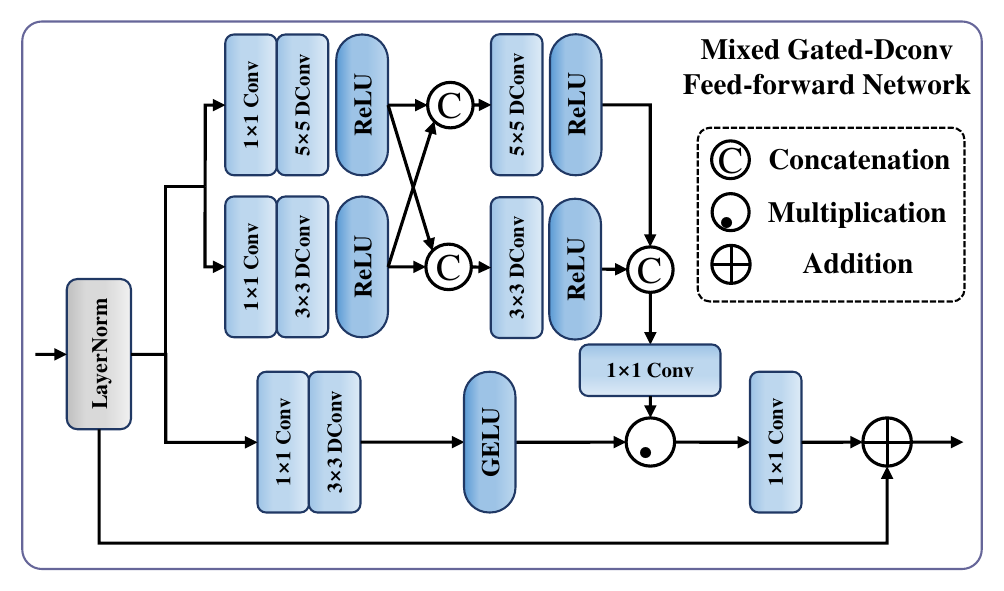}
\caption{The structure diagram of the Mixed Gated-Dconv Feed-forward Network.}
\label{fig9}
\end{figure}

\noindent\textbf{Decoder Architecture.} 
In accordance with Fig.~\ref{fig4} and Fig.~\ref{fig7}, each decoder block contains $N_i^{dec}$ reconstruction modules, which are composed alternately of SGS-Transformer and Dense Transformer. The Transformer paradigm applied in the reconstruction modules alternately focuses on global and local information, demonstrating a more comprehensive receptive field, as shown in Fig.~\ref{fig6} (d). Residual connections are employed in each reconstruction module. The procedures for reconstructing within a reconstruction module can be defined as
\begin{equation}
    \boldsymbol{D_i^0} = \mathcal{SA}\left(\ell_n(\boldsymbol{D_{i+1}}),\boldsymbol{S_i}\right)+\boldsymbol{D_{i-1}},
\label{eq12}
\end{equation}
\begin{equation}
    \boldsymbol{D_i^1}=f_n\left(\ell_n(\boldsymbol{D_i^0})\right) + \boldsymbol{D_{i-1}},
\label{eq13}
\end{equation}
\begin{equation}
    \boldsymbol{D_i^2} = \mathcal{DA}(\ell_n(\boldsymbol{D_i^1}),\boldsymbol{D_i^1})+\boldsymbol{D_i^1},
\label{eq14}
\end{equation}
\begin{equation}
    \boldsymbol{D_i}=f_n\left(\ell_n(\boldsymbol{D_i^2})\right) + \boldsymbol{D_i^1},
\label{eq15}
\end{equation}
where $\mathcal{SA}(\cdot)$ represents the SGSA, $\mathcal{DA}(\cdot)$ is the dense attention, $\ell_n$ denotes the layer normalization and $f_n$ denotes the feed-forward network. $\boldsymbol{D_i^n}$ represents the output from the $n$-th reconstructing module of the $i$-th decoder, and $\boldsymbol{D_i}$ is the final output of the $i$-th decoder. 

\begin{figure*}[t]
\centering
\includegraphics[width=0.95\textwidth]{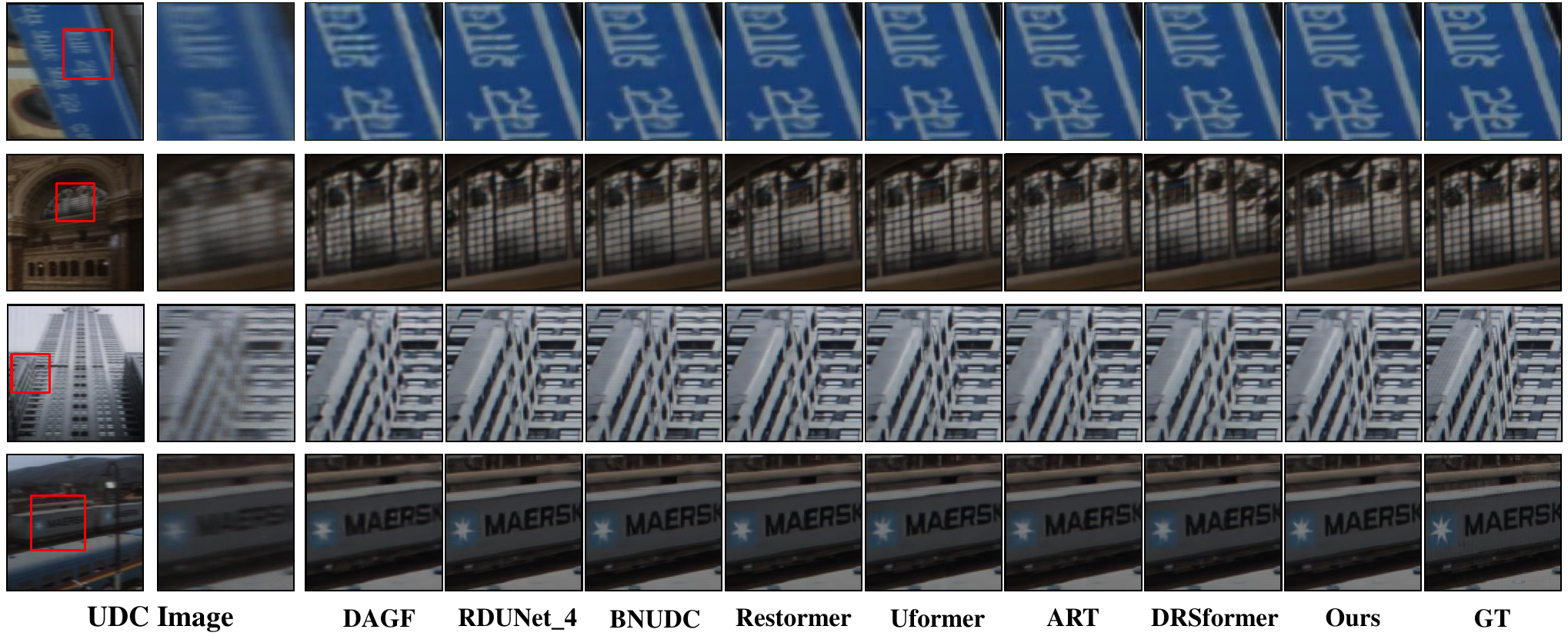}
\caption{Visual comparison results of the T-OLED dataset. The first column is the UDC degraded images, with the red-boxed region representing the selected area for comparison. Subsequent columns show the comparison results and the ground truth.}
\label{fig10}
\end{figure*}

\begin{table*}[t]
\caption{Quantitative comparison on t-OLED dataset. The best and the second-best results are indicated in red and blue respectively.
}
\centering
\begin{tabular}{c|c|c|cccc|cccc}
\hline
 & \multirow{2}{*}{Method} & \multirow{2}{*}{\# PARAM} & \multicolumn{4}{c|}{TEST SET} & \multicolumn{4}{c}{VALIDATION SET} \\
 &  &  & PSNR$\uparrow$ & SSIM$\uparrow$ & LPIPS$\downarrow$ & DISTS$\downarrow$ & PSNR$\uparrow$ & SSIM$\uparrow$ & LPIPS$\downarrow$ & DISTS$\downarrow$ \\ \cline{2-11} 
 & INPUT & - & 28.44 & 0.9102 & 0.3637 & 0.3180 & 29.80 & 0.9179 & 0.3573 & 0.3116 \\ \hline
\multirow{4}{*}{\begin{tabular}[c]{@{}c@{}}UDC\\Methods\end{tabular}}
 & MSUNET~\cite{UDCIR}& 8.9M & 37.40 & 0.9756 & 0.1093 & 0.1052 & 38.25 & 0.9772 & 0.1174 & 0.1155 \\
 & DAGF~\cite{DAGF}& 1.1M & 36.59 & 0.9724 & 0.1416 & 0.1254 & 37.50 & 0.9738 & 0.1506 & 0.1376 \\
 & RDUNet\_4~\cite{udcresnet}& 31.7M & 38.21 & 0.9800 & 0.0988 & 0.0964 & 39.01 & 0.9813 & \textcolor{blue}{0.1040} & \textcolor{blue}{0.1002} \\
 & BNUDC~\cite{BNUDC}& 4.6M & 38.22 & 0.9798 & 0.1748 & 0.1511 & \textcolor{blue}{39.09} & 0.9814 & 0.1072 & 0.1052 \\ \hline
\multirow{2}{*}{\begin{tabular}[c]{@{}c@{}}Transformer\\Methods\end{tabular}} & Restormer~\cite{restormer}& 26.13M & 37.07 & 0.9783 & 0.0982 & \textcolor{blue}{0.0919} & 37.59 & 0.9793 & 0.1075 & 0.1023 \\
 & Uformer~\cite{uformer}& 20.63M & \textcolor{blue}{38.33} & \textcolor{blue}{0.9802} & \textcolor{blue}{0.0967} & 0.0936 & 39.03 & \textcolor{blue}{0.9815} & 0.1056 & 0.1036 \\ \hline
\multirow{3}{*}{\begin{tabular}[c]{@{}c@{}}Sparse\\Transformer\\Methods\end{tabular}} & ART~\cite{sparse11}& 16.15M & 37.01 & 0.9775 & 0.1005 & 0.0951 & 37.60 & 0.9789 & 0.1090 & 0.1049 \\
 & DRSformer~\cite{sparse12}& 23.78M & 37.60 & 0.9791 & 0.1020 & 0.0967 & 38.19 & 0.9802 & 0.1107 & 0.1063 \\ \cline{2-11} 
 & Proposed (Ours) & 9.3M & \textcolor{red}{38.42} & \textcolor{red}{0.9806} & \textcolor{red}{0.0926} & \textcolor{red}{0.0883} & \textcolor{red}{39.10} & \textcolor{red}{0.9818} & \textcolor{red}{0.1029} & \textcolor{red}{0.0993} \\ \hline 
\end{tabular}
\label{table_T}
\end{table*}

\subsection{Encoder Block}
\label{sec:encoder}
Encoder blocks with multi-scale are responsible for transforming pixel-level information in images into high-level features. 
Same as decoders, previous studies~\cite{restormer,uformer} also achieve remarkable success by employing Vision Transformer in encoder blocks for the extraction of multi-scale deep features. 
Likewise, global self-attention encoding, while bringing about more comprehensive contextual features, also introduces a substantial amount of redundant information and noise~\cite{sparse12,sparse13}.
Similar to decoders, to mitigate the adverse effects of redundancy information and noise, we replace global self-attention with sparse self-attention.
But unlike decoders, we employ the lightweight version of SGS-Transformer in encoders, which removes the Dense Transformer to reduce unnecessary sampling of global information in feature encoding. In addition, considering that excessive guiding information may impose constraints on the encoding process, it also removes the instance segmentation feature from sparse attention's key matrix. The detailed information and structure of the lightweight version of SGS-Transformer and encoder block are elaborated in the following.

\noindent\textbf{Light Segmentation Guided Sparse Attention.}
\label{lsgsa}
Unlike the reconstructing process of decoders, considering that there is no need to sample the extensive global information exhaustively during encoding, we remove Dense Transformers to enhance the efficiency of the encoding process. In addition, in the SGS-Transformer, we employ a Segmentation Guided Sparse Attention (SGSA) that differs from the one used in decoders. Specifically, for the encoding process, the SGSA in decoders would impose overly strong guidance on the attention. Therefore, we remove the introduction of the instance segmentation feature into sparse attention's Key matrix to avoid imposing constraints on the encoding process. This simplified SGSA is named as the lightweight version of SGSA (l-SGSA), as illustrated in Fig.~\ref{fig8}. The l-SGSA can be defined as
\begin{equation}
    \mathcal{SA}_l(\boldsymbol{Q,K,V,S_i})=softmax\left[\mathcal{T}_k\left(\frac{\boldsymbol{QK}^T}{\sqrt{\lambda}}\right)\right](\boldsymbol{V}\odot \boldsymbol{S_i}),
\label{eq16}
\end{equation}
where $\boldsymbol{E_i^n}$ represents the input, and $\mathcal{T}_k(\cdot)$ denotes the top-k ranking strategy. $\boldsymbol{S_i}$ is the instance segmentation features map. $\odot$ denotes the element-wise multiplication.

\noindent\textbf{Channel Attention Activate Block.} 
\label{CAAB}
The channel attention aims to assign different levels of importance to different channels, enabling the model to emphasize the most important features and suppress less informative channels. This aligns with the use of sparse attention. Therefore, before employing l-SGSA, we introduce the Channel Attention Activate Block (CAAB) to activate features and enhance the encoding efficiency of l-SGSA. They respectively correspond to the channel-wise features and pixel-wise features.

\begin{figure*}[t]
\centering
\includegraphics[width=0.95\textwidth]{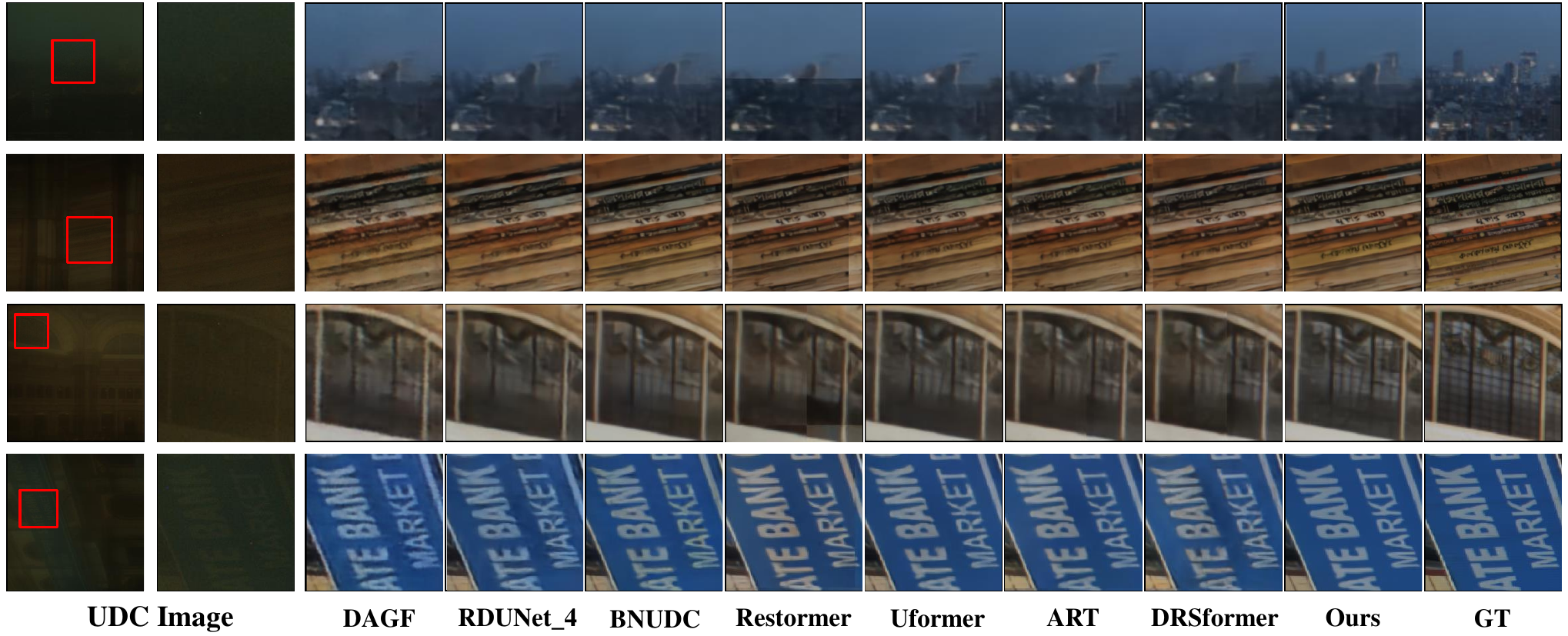}
\caption{Visual comparison results on the P-OLED dataset. The first column is the UDC degraded images, with the red-boxed region representing the selected area for comparison. Subsequent columns show the comparison results and the ground truth.}
\label{fig11}
\end{figure*}

\begin{table*}[t]
\caption{Quantitative comparison on P-OLED dataset. The best and the second-best results are indicated in red and blue respectively.}
\centering
\begin{tabular}{c|c|c|cccc|cccc}
\hline
 & \multirow{2}{*}{Method} & \multirow{2}{*}{\# PARAM} & \multicolumn{4}{c|}{TEST SET} & \multicolumn{4}{c}{VALIDATION SET} \\
 &  &  & PSNR$\uparrow$ & SSIM$\uparrow$ & LPIPS$\downarrow$ & DISTS$\downarrow$ & PSNR$\uparrow$ & SSIM$\uparrow$ & LPIPS$\downarrow$ & DISTS$\downarrow$ \\ \cline{2-11} 
 & INPUT & - & 15.77 & 0.6805 & 0.7704 & 0.5701 & 16.34 & 0.7071 & 0.7009 & 0.5685 \\ \hline
\multirow{4}{*}{\begin{tabular}[c]{@{}c@{}}UDC\\Methods\end{tabular}} & MSUNET~\cite{UDCIR}& 8.9M & 29.17 & 0.9393 & 0.2239 & 0.1746 & 29.96 & 0.9343 & 0.2281 & 0.1774 \\
 & DAGF~\cite{DAGF}& 1.1M & 31.99 & 0.9492 & 0.2309 & 0.2042 & 33.28 & 0.9545 & 0.2385 & 0.2064 \\
 & ResUNet\_5~\cite{udcresnet}& 82.21M & 31.40 & 0.9509 & 0.2146 & 0.1856 & 32.54 & 0.9541 & 0.2229 & 0.1951 \\
 & BNUDC~\cite{BNUDC}& 4.6M & \textcolor{red}{33.39} & \textcolor{blue}{0.9610} & \textcolor{blue}{0.1748} &0.1511  & \textcolor{red}{34.39} & \textcolor{blue}{0.9634} & \textcolor{blue}{0.1871} & 0.1612 \\ \hline
\multirow{2}{*}{\begin{tabular}[c]{@{}c@{}}Transformer\\Methods\end{tabular}} & Restormer~\cite{restormer}& 26.13M & 28.87 & 0.9402 & 0.2053 & \textcolor{blue}{0.1511} & 29.53 & 0.9423 & 0.2168 & \textcolor{blue}{0.1609} \\
 & Uformer~\cite{uformer}& 20.63M & 32.32 & 0.9569 & 0.1860 & 0.1623 & 33.23 & 0.9594 & 0.1965 & 0.1710 \\ \hline
\multirow{3}{*}{\begin{tabular}[c]{@{}c@{}}Sparse\\Transformer\\Methods\end{tabular}} & ART~\cite{sparse11}& 16.15M & 31.11 & 0.9518 & 0.1946 & 0.1608 & 31.70 & 0.9528 & 0.2064 & 0.1688 \\
 & DRSformer~\cite{sparse12}& 23.78M & 30.14 & 0.9478 & 0.2034 & 0.1625 & 30.97 & 0.9489 & 0.2110 & 0.1695 \\ \cline{2-11} 
 & Proposed (Ours) & 9.3M & \textcolor{blue}{33.28} & \textcolor{red}{0.9627} & \textcolor{red}{0.1573} & \textcolor{red}{0.1286} & \textcolor{blue}{33.60} & \textcolor{red}{0.9639} & \textcolor{red}{0.1744} & \textcolor{red}{0.1443} \\ \hline 
\end{tabular}
\label{table_P}
\end{table*}

\noindent\textbf{Mixed Gated-Dconv Feed-forward Network.} 
\label{MGFN}
In the conclusions of prior studies~\cite{transformer,vit,sparse12,restormer}, strengthening the perception ability of feed-forward networks is considered a pivotal factor in improving the performance of Vision Transformer. Due to the fewer number of Transformers in the encoders compared to the decoders, such a modification does not lead to a substantial escalation in the number of parameters.
Inspired by \cite{sparse12}, we introduce multi-scale mixed structure into a Gated-Dconv Feed-forward Network, and the mixing of multi-scale features enables the feed-forward network to learn more local structural information for the reconstruction of degraded image regions. Specifically, as illustrated in Fig.~\ref{fig9}, we replace the single-scale depth-wise $3\times3$ convolution in the gating branch with a combination of $3\times3$ convolution and $5\times5$ convolution, and concatenate them after the ReLU activation function. Subsequently, after reducing channels through a $1\times1$ convolution, the resulting feature of the gating branch maps is element-wise multiplied by the feature maps from the main branch. Given an input tensor $\boldsymbol{\mathcal{X}}\in\mathbb{R}^{H\times W\times C}$, MGFN is formulated as
\begin{equation}
\begin{split}
&\boldsymbol{\hat{\mathcal{X}}}=f^c_{1\times1}\left(Ma(\boldsymbol{\mathcal{X}})\odot Ga(\boldsymbol{\mathcal{X}})\right)+\boldsymbol{\mathcal{X}},\\
&Ma(\boldsymbol{\mathcal{X}})=\tau\left(f^{dwc}_{3\times3}\left(f^{conv}_{1\times1}(\boldsymbol{\mathcal{X}})\right)\right),\\
&\boldsymbol{\mathcal{X}_1}=\tau\left(f^{dwc}_{3\times3}\left(f^{conv}_{1\times1}(\boldsymbol{\mathcal{X}})\right)\right),\\
&\boldsymbol{\mathcal{X}_2}=\tau\left(f^{dwc}_{3\times3}\left(f^{conv}_{1\times1}(\boldsymbol{\mathcal{X}})\right)\right),\\
&Ga(\boldsymbol{\mathcal{X}})=f^c_{1\times1}\left(\tau\left(f^{dwc}_{3\times3}(\boldsymbol{\mathcal{X}_1}\circ\boldsymbol{\mathcal{X}_2})\right)\circ\tau\left(f^{dwc}_{5\times5}(\boldsymbol{\mathcal{X}_2}\circ\boldsymbol{\mathcal{X}_1})\right)\right),\\
\end{split}
\label{eqmgfn}
\end{equation}
where $\odot$ is the element-wise multiplication, and $\circ$ is the channel-wise concatenation. $\tau$ is an activation function, representing ReLU or GELU. $f^{conv}_{1\times1}$ denotes $1\times1$ convolution, while $f^{dwc}_{i\times i}$ denotes $i\times i$ depth-wise convolution. $Ma(\boldsymbol{\mathcal{X}})$ is the Main-Branch of the Gated-Dconv network, while $Ga(\boldsymbol{\mathcal{X}})$ is the Gating-Branch of the Gated-Dconv network.

\noindent\textbf{Encoder Architecture.} Fig.~\ref{fig8} shows that there are $N_i^{enc}$ light SGS-Transformer and a Channel Attention Activate Block (CAAB) in the $i$-th encoder block. For each encoder block, initially inputted is the multi-scale feature tensor $\boldsymbol{E_{i-1}}$. Notably, before the $\boldsymbol{I_{udc}}$ enters the first encoder, we initially extract the feature tensor $\boldsymbol{E_0}\in\mathbb{R}^{H\times W\times C}$ through UDC Shallow Feature Extractor (USFE). And then, $\boldsymbol{E_{i-1}}$ is activated by CAAB before entering the Transformers, aiming to enhance the encoding efficiency of subsequent blocks. Subsequently, using the segmentation feature matrices $\boldsymbol{S_i}$ as guided signal, consecutive light SGS-Transformers perform encoding to extract high-level semantic features, resulting in the final output, the multi-scale feature tensor $\boldsymbol{E_i}$. The aforementioned encoding process can be formulated as
\begin{equation*}
    \boldsymbol{E_i^0} = \mathcal{CA}(\boldsymbol{E_{i-1}}),
\label{eq17.5}
\end{equation*}
\begin{equation}
    \boldsymbol{\Bar{E}_i^n} = \mathcal{SA}_l\left(\ell_n(\boldsymbol{E_i^{n-1}}),S_i\right)+\boldsymbol{E_i^{n-1}},
\label{eq17}
\end{equation}
\begin{equation}
    \boldsymbol{E_i^n}=f_n\left(\ell_n(\boldsymbol{\Bar{E}_i^n})\right) + \boldsymbol{E_i^{n-1}},
\label{eq18}
\end{equation}
where $\mathcal{CA}(\cdot)$ is the CAAB, $\mathcal{SA}_l(\cdot)$ represents l-SGSA, $\ell_n$ denotes the layer normalization and $f_n$ denotes the MGFN. $\boldsymbol{E_i^n}$ represents the output from the $n$-th light SGS-Transformer of the $i$-th encoder, and $\boldsymbol{\Bar{E}_i^n}$ is the output from l-SGSA. 

\subsection{Loss Function}
\label{sec:loss}
We incorporate a perceptual loss term to enhance the model's ability by acquiring richer and more comprehensive feature representations. Thus the loss function is
\begin{equation}
\begin{split}
\hat{\mathcal{L}}(\boldsymbol{I_{r},I_{gt}})&=\lambda_1\mathcal{L}_{1}(\boldsymbol{I_{r},I_{gt}})+\lambda_4\mathcal{L}_{perceptual}(\boldsymbol{I_{r},I_{gt}})\\
&+\lambda_2\mathcal{L}_{psnr}(\boldsymbol{I_{r},I_{gt}})+\lambda_3\mathcal{L}_{ssim}(\boldsymbol{I_{r},I_{gt}}),
\end{split}
\label{eqloss}
\end{equation}
where $\boldsymbol{I_{r}}$ is the UDC degraded image and $\boldsymbol{I_{gt}}$ is the ground-truth. $\mathcal{L}_{1}$, $\mathcal{L}_{perceptual}$, $\mathcal{L}_{psnr}$ and $\mathcal{L}_{ssim}$ represent L1 loss, perceptual loss~\cite{perceptualloss}, PSNR loss~\cite{psnrloss}, and SSIM loss~\cite{ssimloss} respectively. [$\mathcal{L}_{1}$, $\mathcal{L}_{2}$, $\mathcal{L}_{3}$, $\mathcal{L}_{4}$] is set to [$1$, $0.2$, $0.2$, $1$] in the early stage of training and [$0$, $0.2$, $0.1$, $1$] in the later stage of training.

\begin{table*}[t]
\caption{Ablation studies on the designed modules of our SGSFormer.}
\centering
\begin{tabular}{c|c|ccc|ccc|cccc}
\hline
 & Seg Module & \multicolumn{3}{c|}{Attention Type} & \multicolumn{3}{c|}{Encoder Component} & \multicolumn{4}{c}{T-OLED TEST SET} \\ \hline
Variant & Seg Guid & S+D & Sparse & Dense & l-SGST & CAAB & FN  & PSNR$\uparrow$ & SSIM$\uparrow$ & LPIPS$\downarrow$ & DISTS$\downarrow$ \\ \hline
w/o Segmentation Guidance &  & \checkmark &  &  & \checkmark & \checkmark & MGFN & 38.05 & 0.9792 & 0.0964 & 0.0927 \\
w/o Dense Attention & \checkmark &  & \checkmark &  & \checkmark & \checkmark & MGFN & 38.24 & 0.9802 & 0.0946 & 0.0914 \\
v/o Sparse Attention & \checkmark &  &  & \checkmark & \checkmark & \checkmark & MGFN & 38.11 & 0.9799 & 0.0981 & 0.0933 \\
Use full SGS-Transformer & \checkmark & \checkmark &  &  &  & \checkmark & MGFN & 37.05 & 0.9780 & 0.1148 & 0.1056 \\
w/o CAAB & \checkmark & \checkmark &  &  & \checkmark &  & MGFN & 37.84 & 0.9793 & 0.0937 & 0.0869 \\
Replace MGFN with MFFN & \checkmark & \checkmark &  &  & \checkmark & \checkmark & MFFN & 38.19 & 0.9803 & 0.0943 & 0.0871 \\
Replace MGFN with GDFN\cite{restormer} & \checkmark & \checkmark &  &  & \checkmark & \checkmark & GDFN & 38.12 & 0.9799 & 0.0936 & 0.0883 \\
Replace MGFN with MSFN\cite{sparse12} & \checkmark & \checkmark &  &  & \checkmark & \checkmark & MSFN & 38.29 & 0.9765 & 0.0930 & 0.0891 \\ \hline
Proposed (Ours) & \checkmark & \checkmark &  &  & \checkmark & \checkmark & MGFN & 38.42 & 0.9806 &0.0926 & 0.0883 \\ \hline
\end{tabular}
\label{Ablation}
\end{table*}

\begin{table}[t]
\caption{Ablation studies on the fusion method of instance segmentation features.}
\centering
\begin{tabular}{c|cc}
\hline
Fusion Method & PSNR$\uparrow$ & SSIM$\uparrow$ \\ \hline
$1\times1$ Conv & 38.23 & 0.9793 \\
Linear & 38.16 & 0.9789 \\
Addition & 38.31 & 0.9801 \\
Dot product & 37.74 & 0.9776  \\
\hline
Multiplication & 38.42 & 0.9806 \\ \hline
\end{tabular}
\label{Ablation2}
\end{table}

\section{Experiment}
\label{sec:experiment}

\subsection{Implementation Details}
The proposed network structure has four layers of encoder blocks, one layer of latent blocks, and four layers of decoder blocks. Each layer of the encoder block consists of \{4, 6, 7, 8\} Channel-Attention Activate Blocks, and \{1, 1, 1, 1\} lightweight SGS-Transformers, with the numbers of heads being \{1, 2, 4, 8\}. The latent layer comprises 8 transformer blocks with 16 multi-heads. For each encoder block layer, the numbers of reconstruction modules, composed alternately of SGS-Transformer and Dense Transformer, are \{2, 3, 3, 4\}, with the corresponding numbers of heads being \{1, 2, 4, 8\}.

All networks use the Adam optimizer~\cite{adam} with $\beta_1$ set to $0.9$ and $\beta_2$ set to $0.999$, and cyclical learning rate~\cite{smith2017cyclical} with periodic variations between $1 \times 10^{-4}$ and $1 \times 10^{-5}$. We utilize horizontal flipping and rotation as data augmentation, along with cropping image patches sized at $256 \times 256$ for training. The batch size is set to 24. We use the P-OLED dataset preprocessed by Koh\etal\cite{BNUDC} for training. The network for T-OLED dataset is trained at about $6\times10^4$ epochs, while the network for P-OLED dataset is trained about $1\times10^5$ epochs.

\subsection{Comparisons with SOTA Methods}
We conduct comparisons on two benchmark datasets: T-OLED and P-OLED~\cite{UDCIR}. And we use four evaluation metrics: PSNR for measuring pixel-wise distance; SSIM for measuring structural similarity~\cite{ssim}; LPIPS and DISTS for measuring perceptual similarity~\cite{lpips,dists}. All the aforementioned metrics are measured on $float32$ matrices within the value range [0, 1]. 

We select the methods for comparison from three domains. Firstly, we select the previous UDC image restoration methods, which mostly employ CNNs, including MSUNET~\cite{UDCIR}, BNUDC~\cite{BNUDC}, DAGF~\cite{DAGF}, RDUNet and ResUNet~\cite{udcresnet}. Secondly, we select classic Transformer networks for other image restoration tasks, including Restormer~\cite{restormer} and Uformer~\cite{uformer}. Lastly, we select several Vision sparse Transformer networks: ART~\cite{sparse11} and DRSformer~\cite{sparse12}. All models for comparison are trained until convergence.

\noindent\textbf{T-OLED Dataset.} Table \ref{table_T} presents the quantitative comparison experimental results of the aforementioned methods on the T-OLED dataset. Our SGSFormer achieves state-of-the-art performance regarding all four evaluation metrics, particularly in terms of perceptual similarity: LPIPS and DISTS. In addition, it can be observed that the Transformer methods generally outperform the CNN methods in terms of perceptual similarity, as measured by LPIPS and DISTS. This indicates that the global receptive field of the Transformer methods plays a significantly enhanced role in improving the perceptual similarity of UDC image restoration. However, the Transformer methods do not demonstrate an advantage over CNN methods in terms of PSNR and SSIM. This indicates that the Transformer method has a deficiency in optimizing pixel-wise distance and texture information. Specifically, redundant information and noise introduced by global attention impair the performance of pixel-level restoration. Our SGSFormer mitigates this deficiency, and the results demonstrate significant improvements in both PSNR and SSIM compared to other Transformer methods. The visual comparison results are shown in Fig.~\ref{fig10}. Our SGSFormer achieves clearer and less smearing in enlarged area details, especially in the restoration of text contours and texture details.

\noindent\textbf{P-OLED Dataset.} Table \ref{table_P} presents the quantitative comparison experimental results of the aforementioned methods on the P-OLED dataset. Our SGSFormer outperforms others in state-of-the-art performance on three metrics: SSIM, LPIPS, and DISTS, and ranks as the second best on the PSNR metric. Similar to the results on the T-OLED dataset, we can observe that the Transformer methods generally outperform the CNN methods in terms of perceptual similarity. However, the performance of Transformer methods on PSNR and SSIM has deteriorated compared to that on the T-OLED dataset. This deterioration can be attributed to the presence of more redundant information and noise in the degraded images of the P-OLED dataset compared to those in the T-OLED dataset, thus amplifying the deficiency of the Transformer method. Consistent with the results of the T-OLED dataset, the experimental results of the P-OLED dataset indicate that our SGSFormer alleviates this deficiency, with significant improvements in both PSNR and SSIM compared to other Transformer methods. The visual comparison results are shown in Fig.~\ref{fig11}. Our SGSFormer clearly excels in recovering finer details and achieving more accurate color restoration. Its restored images exhibit clearer textures with less smudging.

\subsection{Ablation Studies}
We conduct ablation experiments on all modules we design to validate their effectiveness. Considering factors related to the convergence speed of models, all ablation experiments are conducted on the T-OLED dataset and validated on the test set. All ablation models are trained until convergence. The quantitative results of the ablation experiments are shown in Table~\ref{Ablation}.

\subsubsection{Instance Segmentation Guidance}
To validate the enhancing effect of instance segmentation features on the top-k ranking strategy within sparse attention, we conduct an ablation experiment by excluding instance segmentation features from all sparse attention while retaining all other factors unchanged. The results in Table~\ref{Ablation} demonstrate a decline in performance across all four evaluation metrics. This indicates that the introduction of instance segmentation features has a positive impact on sparse attention. Table~\ref{Ablation2} presents the ablation experiments of the fusion method, which integrates instance segmentation features into sparse attention. The results indicate that the most significant improvement is achieved through element-wise multiplication.

\subsubsection{Alternating Application of Sparse and Dense Attention}
We conduct two ablation experiments to validate whether the combined effect of sparse and dense attention would yield superior results: replacing all sparse attention with dense attention, and conversely, replacing all dense attention with sparse attention. The results in Table~\ref{Ablation} indicate that using solely dense attention or sparse attention alone would result in performance degradation. Additionally, sparse attention alone performs marginally better than the case of dense attention alone.
\subsubsection{Light Version of SGS-Transformer in Encoders}
We adopt the lightweight version of SGS-Transformer in encoders, instead of the full version (Sec. \ref{lsgsa}). This decision is based on the concern that introducing excessive segmentation guidance may impose constraints on the encoding process, thereby impacting the performance of the model. The results in Table ~\ref{Ablation} validate this perspective.
\subsubsection{Channel-Attention Activate Block}
Sparse attention operates on pixel-wise features, while channel attention operates on channel-wise features. These two modules complement each other effectively (Sec. \ref{CAAB}). The results in Table \ref{Ablation} demonstrate that CAAB has a positive impact on the encoding process.
\subsubsection{Mixed Gated-Dconv Feed-forward Network}
We compare MGFN with three baseline feed-forward networks to validate its efficacy (Sec. \ref{MGFN}). They are Multi-Feature Feed-forward Network (MFFN), Gated-Dconv Feed-forward Network) (GDFN)~\cite{restormer} and Mixed-Scale Feed-forward Network (MSFN)~\cite{sparse12}. Table \ref{Ablation} illustrates that our MGFN achieves the best performance.

\section{Conclusion and Future Work}
\label{sec:conclusion}
In this work, we propose the introduction of Instance Segmentation Maps as guiding information to enhance the performance of sparse attention. And devise a paradigm consisting of alternating Segmentation-Guided Sparse Attention and Dense Attention, enabling the simultaneous consideration of globally generalized information and locally crucial details for image reconstruction. 
Based on the aforementioned propositions, we design an asymmetric U-net network called SGSFormer. Numerous experiments demonstrate that our SGSFormer achieves state-of-the-art performance in restoring UDC degraded images.

As an exploratory work, our study has certain limitations and areas for improvement. First, we fine-tune an instance segmentation network with strong zero-shot performance instead of training a dedicated instance segmentation network specifically tailored for UDC degraded images. This could render the instance segmentation module a bottleneck for overall network performance. Second, similar to most Transformer networks, our SGSFormer exhibits the drawbacks of a high parameter count and significant computational overhead.
In the future, we will try to address the aforementioned shortcomings, continually refine our approach, and achieve superior performance. Furthermore, we will continue exploring and extending the Segmentation Guided Sparse Transformer in various low-level vision tasks.

\bibliographystyle{IEEEtran}
\bibliography{myreference}

\begin{thebibliography}{10}
\providecommand{\url}[1]{#1}
\csname url@samestyle\endcsname
\providecommand{\newblock}{\relax}
\providecommand{\bibinfo}[2]{#2}
\providecommand{\BIBentrySTDinterwordspacing}{\spaceskip=0pt\relax}
\providecommand{\BIBentryALTinterwordstretchfactor}{4}
\providecommand{\BIBentryALTinterwordspacing}{\spaceskip=\fontdimen2\font plus
\BIBentryALTinterwordstretchfactor\fontdimen3\font minus \fontdimen4\font\relax}
\providecommand{\BIBforeignlanguage}[2]{{%
\expandafter\ifx\csname l@#1\endcsname\relax
\typeout{** WARNING: IEEEtran.bst: No hyphenation pattern has been}%
\typeout{** loaded for the language `#1'. Using the pattern for}%
\typeout{** the default language instead.}%
\else
\language=\csname l@#1\endcsname
\fi
#2}}
\providecommand{\BIBdecl}{\relax}
\BIBdecl

\bibitem{yanshe}
Z.~Qin, Y.-H. Tsai, Y.-W. Yeh, Y.-P. Huang, and H.-P.~D. Shieh, ``See-through image blurring of transparent organic light-emitting diodes display: calculation method based on diffraction and analysis of pixel structures,'' \emph{Journal of Display Technology}, pp. 1242--1249, 2016.

\bibitem{UDCIR}
Y.~Zhou, D.~Ren, N.~Emerton, S.~Lim, and T.~Large, ``Image restoration for under-display camera,'' in \emph{IEEE Conference on Computer Vision and Pattern Recognition}, 2021, pp. 9179--9188.

\bibitem{DISCNet}
R.~Feng, C.~Li, H.~Chen, S.~Li, C.~C. Loy, and J.~Gu, ``Removing diffraction image artifacts in under-display camera via dynamic skip connection network,'' in \emph{IEEE Conference on Computer Vision and Pattern Recognition}, 2021, pp. 662--671.

\bibitem{pdl}
A.~Yang and A.~C. Sankaranarayanan, ``Designing display pixel layouts for under-panel cameras,'' \emph{IEEE Transactions on Pattern Analysis and Machine Intelligence}, 2021.

\bibitem{pdl2}
Z.~Zhang, ``14.4: Diffraction simulation of camera under display,'' in \emph{SID Symposium Digest of Technical Papers}, 2021, pp. 93--96.

\bibitem{pdl3}
C.~X. Xu, Q.~Yao, W.~He, S.~Shu, and G.~C. Yuan, ``P-125: A novel method to increase the transmittance of full display with camera,'' in \emph{SID Symposium Digest of Technical Papers}, 2023, pp. 1316--1318.

\bibitem{csvt1}
Y.~Li, J.~Wu, and Z.~Shi, ``Lightweight neural network for enhancing imaging performance of under-display camera,'' \emph{IEEE Transactions on Circuits and Systems for Video Technology}, 2023.

\bibitem{kwon2021controllable}
K.~Kwon, E.~Kang, S.~Lee, S.-J. Lee, H.-E. Lee, B.~Yoo, and J.-J. Han, ``Controllable image restoration for under-display camera in smartphones,'' in \emph{IEEE Conference on Computer Vision and Pattern Recognition}, 2021, pp. 2073--2082.

\bibitem{DAGF}
V.~Sundar, S.~Hegde, D.~Kothandaraman, and K.~Mitra, ``Deep atrous guided filter for image restoration in under display cameras,'' in \emph{European Conference on Computer Vision Workshop}, 2020, pp. 379--397.

\bibitem{mipi}
R.~Feng, C.~Li, S.~Zhou, W.~Sun, Q.~Zhu, J.~Jiang, Q.~Yang, C.~C. Loy, J.~Gu, Y.~Zhu \emph{et~al.}, ``Mipi 2022 challenge on under-display camera image restoration: Methods and results,'' in \emph{European Conference on Computer Vision Workshop}, 2023, pp. 60--77.

\bibitem{udcresnet}
Q.~Yang, Y.~Liu, J.~Tang, and T.~Ku, ``Residual and dense unet for under-display camera restoration,'' in \emph{European Conference on Computer Vision Workshop}, 2021, pp. 398--408.

\bibitem{PDCRN}
H.~Panikkasseril~Sethumadhavan, D.~Puthussery, M.~Kuriakose, and J.~Charangatt~Victor, ``Transform domain pyramidal dilated convolution networks for restoration of under display camera images,'' in \emph{European Conference on Computer Vision Workshop}, 2020, pp. 364--378.

\bibitem{ECCVchallenge}
Y.~Zhou, M.~Kwan, K.~Tolentino, N.~Emerton, S.~Lim, T.~Large, L.~Fu, Z.~Pan, B.~Li, Q.~Yang \emph{et~al.}, ``Udc 2020 challenge on image restoration of under-display camera: Methods and results,'' in \emph{European Conference on Computer Vision Workshop}, 2020, pp. 337--351.

\bibitem{UDCUNet}
X.~Liu, J.~Hu, X.~Chen, and C.~Dong, ``Udc-unet: Under-display camera image restoration via u-shape dynamic network,'' in \emph{European Conference on Computer Vision Workshop}, 2023, pp. 113--129.

\bibitem{qi2021isp}
M.~Qi, Y.~Li, and W.~Heidrich, ``Isp-agnostic image reconstruction for under-display cameras,'' \emph{arXiv preprint arXiv:2111.01511}, 2021.

\bibitem{Nonaligned}
R.~Feng, C.~Li, H.~Chen, S.~Li, J.~Gu, and C.~C. Loy, ``Generating aligned pseudo-supervision from non-aligned data for image restoration in under-display camera,'' in \emph{IEEE Conference on Computer Vision and Pattern Recognition}, 2023, pp. 5013--5022.

\bibitem{wacv1}
M.~V. Conde, F.~Vasluianu, J.~Vazquez-Corral, and R.~Timofte, ``Perceptual image enhancement for smartphone real-time applications,'' in \emph{Proceedings of the IEEE/CVF Winter Conference on Applications of Computer Vision}, 2023, pp. 1848--1858.

\bibitem{tjf}
J.~Tan, X.~Chen, T.~Wang, K.~Zhang, W.~Luo, and X.~Cao, ``Blind face restoration for under-display camera via dictionary guided transformer,'' \emph{IEEE Transactions on Circuits and Systems for Video Technology}, 2023.

\bibitem{mix4}
J.-h. Lou, L.~Zhang, and L.~Ge, ``7-1: Invited paper: Udc technology for oled display,'' in \emph{SID Symposium Digest of Technical Papers}, 2022, pp. 44--47.

\bibitem{restormer}
S.~W. Zamir, A.~Arora, S.~Khan, M.~Hayat, F.~S. Khan, and M.-H. Yang, ``Restormer: Efficient transformer for high-resolution image restoration,'' in \emph{IEEE Conference on Computer Vision and Pattern Recognition}, 2022, pp. 5728--5739.

\bibitem{uformer}
Z.~Wang, X.~Cun, J.~Bao, W.~Zhou, J.~Liu, and H.~Li, ``Uformer: A general u-shaped transformer for image restoration,'' in \emph{IEEE Conference on Computer Vision and Pattern Recognition}, 2022, pp. 17\,683--17\,693.

\bibitem{chen2021pre}
H.~Chen, Y.~Wang, T.~Guo, C.~Xu, Y.~Deng, Z.~Liu, S.~Ma, C.~Xu, C.~Xu, and W.~Gao, ``Pre-trained image processing transformer,'' in \emph{IEEE Conference on Computer Vision and Pattern Recognition}, 2021, pp. 12\,299--12\,310.

\bibitem{tu2022maxim}
Z.~Tu, H.~Talebi, H.~Zhang, F.~Yang, P.~Milanfar, A.~Bovik, and Y.~Li, ``Maxim: Multi-axis mlp for image processing,'' in \emph{IEEE Conference on Computer Vision and Pattern Recognition}, 2022, pp. 5769--5780.

\bibitem{guo2015efficient}
Q.~Guo, C.~Zhang, Y.~Zhang, and H.~Liu, ``An efficient svd-based method for image denoising,'' \emph{IEEE Transactions on Circuits and Systems for Video Technology}, vol.~26, no.~5, pp. 868--880, 2015.

\bibitem{wt1}
T.~Wang, X.~Zhang, R.~Jiang, L.~Zhao, H.~Chen, and W.~Luo, ``Video deblurring via spatiotemporal pyramid network and adversarial gradient prior,'' \emph{Computer Vision and Image Understanding}, vol. 203, p. 103135, 2021.

\bibitem{wen2020simple}
F.~Wen, R.~Ying, Y.~Liu, P.~Liu, and T.-K. Truong, ``A simple local minimal intensity prior and an improved algorithm for blind image deblurring,'' \emph{IEEE Transactions on Circuits and Systems for Video Technology}, vol.~31, no.~8, pp. 2923--2937, 2020.

\bibitem{wt2}
T.~Wang, K.~Zhang, T.~Shen, W.~Luo, B.~Stenger, and T.~Lu, ``Ultra-high-definition low-light image enhancement: a benchmark and transformer-based method,'' in \emph{AAAI Conference on Artificial Intelligence}, 2023, pp. 2654--2662.

\bibitem{li2021low}
J.~Li, X.~Feng, and Z.~Hua, ``Low-light image enhancement via progressive-recursive network,'' \emph{IEEE Transactions on Circuits and Systems for Video Technology}, vol.~31, no.~11, pp. 4227--4240, 2021.

\bibitem{jiang2020multi}
K.~Jiang, Z.~Wang, P.~Yi, C.~Chen, B.~Huang, Y.~Luo, J.~Ma, and J.~Jiang, ``Multi-scale progressive fusion network for single image deraining,'' in \emph{IEEE Conference on Computer Vision and Pattern Recognition}, 2020, pp. 8346--8355.

\bibitem{zhang2022beyond}
K.~Zhang, W.~Luo, Y.~Yu, W.~Ren, F.~Zhao, C.~Li, L.~Ma, W.~Liu, and H.~Li, ``Beyond monocular deraining: Parallel stereo deraining network via semantic prior,'' \emph{International Journal of Computer Vision}, vol. 130, no.~7, pp. 1754--1769, 2022.

\bibitem{zhang2019image}
H.~Zhang, V.~Sindagi, and V.~M. Patel, ``Image de-raining using a conditional generative adversarial network,'' \emph{IEEE Transactions on Circuits and Systems for Video Technology}, vol.~30, no.~11, pp. 3943--3956, 2019.

\bibitem{jiang2020decomposition}
K.~Jiang, Z.~Wang, P.~Yi, C.~Chen, Z.~Han, T.~Lu, B.~Huang, and J.~Jiang, ``Decomposition makes better rain removal: An improved attention-guided deraining network,'' \emph{IEEE Transactions on Circuits and Systems for Video Technology}, vol.~31, no.~10, pp. 3981--3995, 2020.

\bibitem{qiu2023mb}
Y.~Qiu, K.~Zhang, C.~Wang, W.~Luo, H.~Li, and Z.~Jin, ``Mb-taylorformer: Multi-branch efficient transformer expanded by taylor formula for image dehazing,'' in \emph{IEEE International Conference on Computer Vision}, 2023, pp. 12\,802--12\,813.

\bibitem{zhang2020multi}
X.~Zhang, T.~Wang, W.~Luo, and P.~Huang, ``Multi-level fusion and attention-guided cnn for image dehazing,'' \emph{IEEE Transactions on Circuits and Systems for Video Technology}, vol.~31, no.~11, pp. 4162--4173, 2020.

\bibitem{sparse2}
S.~Jaszczur, A.~Chowdhery, A.~Mohiuddin, L.~Kaiser, W.~Gajewski, H.~Michalewski, and J.~Kanerva, ``Sparse is enough in scaling transformers,'' \emph{Advances in Neural Information Processing Systems}, pp. 9895--9907, 2021.

\bibitem{yang2023designing}
A.~Yang, E.~Kang, H.-E. Lee, and A.~C. Sankaranarayanan, ``Designing phase masks for under-display cameras,'' in \emph{IEEE International Conference on Computer Vision}, 2023, pp. 10\,637--10\,645.

\bibitem{transparent}
G.~Huseynova, J.-H. Lee, Y.~H. Kim, and J.~Lee, ``Transparent organic light-emitting diodes: advances, prospects, and challenges,'' \emph{Advanced Optical Materials}, p. 2002040, 2021.

\bibitem{unet}
O.~Ronneberger, P.~Fischer, and T.~Brox, ``U-net: Convolutional networks for biomedical image segmentation,'' in \emph{Medical Image Computing and Computer-Assisted Intervention--MICCAI 2015: 18th International Conference, Munich, Germany, October 5-9, 2015, Proceedings, Part III 18}, 2015, pp. 234--241.

\bibitem{resnet}
K.~He, X.~Zhang, S.~Ren, and J.~Sun, ``Deep residual learning for image recognition,'' in \emph{IEEE Conference on Computer Vision and Pattern Recognition}, 2016, pp. 770--778.

\bibitem{BNUDC}
J.~Koh, J.~Lee, and S.~Yoon, ``Bnudc: A two-branched deep neural network for restoring images from under-display cameras,'' in \emph{IEEE Conference on Computer Vision and Pattern Recognition}, 2022, pp. 1950--1959.

\bibitem{song2023under}
B.~Song, X.~Chen, S.~Xu, and J.~Zhou, ``Under-display camera image restoration with scattering effect,'' in \emph{IEEE International Conference on Computer Vision}, 2023, pp. 12\,580--12\,589.

\bibitem{liu2023fsi}
C.~Liu, X.~Wang, S.~Li, Y.~Wang, and X.~Qian, ``Fsi: Frequency and spatial interactive learning for image restoration in under-display cameras,'' in \emph{IEEE International Conference on Computer Vision}, 2023, pp. 12\,537--12\,546.

\bibitem{transformer}
A.~Vaswani, N.~Shazeer, N.~Parmar, J.~Uszkoreit, L.~Jones, A.~N. Gomez, {\L}.~Kaiser, and I.~Polosukhin, ``Attention is all you need,'' \emph{Advances in Neural Information Processing Systems}, 2017.

\bibitem{vit}
A.~Dosovitskiy, L.~Beyer, A.~Kolesnikov, D.~Weissenborn, X.~Zhai, T.~Unterthiner, M.~Dehghani, M.~Minderer, G.~Heigold, S.~Gelly \emph{et~al.}, ``An image is worth 16x16 words: Transformers for image recognition at scale,'' \emph{arXiv preprint arXiv:2010.11929}, 2020.

\bibitem{recognition1}
H.~Touvron, M.~Cord, M.~Douze, F.~Massa, A.~Sablayrolles, and H.~J{\'e}gou, ``Training data-efficient image transformers \& distillation through attention,'' in \emph{International Conference on Machine Learning}, 2021, pp. 10\,347--10\,357.

\bibitem{recognition2}
L.~Yuan, Y.~Chen, T.~Wang, W.~Yu, Y.~Shi, Z.-H. Jiang, F.~E. Tay, J.~Feng, and S.~Yan, ``Tokens-to-token vit: Training vision transformers from scratch on imagenet,'' in \emph{IEEE International Conference on Computer Vision}, 2021, pp. 558--567.

\bibitem{swint}
Z.~Liu, Y.~Lin, Y.~Cao, H.~Hu, Y.~Wei, Z.~Zhang, S.~Lin, and B.~Guo, ``Swin transformer: Hierarchical vision transformer using shifted windows,'' in \emph{IEEE International Conference on Computer Vision}, 2021, pp. 10\,012--10\,022.

\bibitem{carion2020end}
N.~Carion, F.~Massa, G.~Synnaeve, N.~Usunier, A.~Kirillov, and S.~Zagoruyko, ``End-to-end object detection with transformers,'' in \emph{European Conference on Computer Vision}, 2020, pp. 213--229.

\bibitem{twins}
X.~Chu, Z.~Tian, Y.~Wang, B.~Zhang, H.~Ren, X.~Wei, H.~Xia, and C.~Shen, ``Twins: Revisiting the design of spatial attention in vision transformers,'' \emph{Advances in Neural Information Processing Systems}, vol.~34, pp. 9355--9366, 2021.

\bibitem{zheng2021rethinking}
S.~Zheng, J.~Lu, H.~Zhao, X.~Zhu, Z.~Luo, Y.~Wang, Y.~Fu, J.~Feng, T.~Xiang, P.~H. Torr \emph{et~al.}, ``Rethinking semantic segmentation from a sequence-to-sequence perspective with transformers,'' in \emph{IEEE Conference on Computer Vision and Pattern Recognition}, 2021, pp. 6881--6890.

\bibitem{wang2021pyramid}
W.~Wang, E.~Xie, X.~Li, D.-P. Fan, K.~Song, D.~Liang, T.~Lu, P.~Luo, and L.~Shao, ``Pyramid vision transformer: A versatile backbone for dense prediction without convolutions,'' in \emph{IEEE International Conference on Computer Vision}, 2021, pp. 568--578.

\bibitem{xie2021segformer}
E.~Xie, W.~Wang, Z.~Yu, A.~Anandkumar, J.~M. Alvarez, and P.~Luo, ``Segformer: Simple and efficient design for semantic segmentation with transformers,'' \emph{Advances in Neural Information Processing Systems}, pp. 12\,077--12\,090, 2021.

\bibitem{survey1}
S.~Khan, M.~Naseer, M.~Hayat, S.~W. Zamir, F.~S. Khan, and M.~Shah, ``Transformers in vision: A survey,'' \emph{ACM computing surveys (CSUR)}, pp. 1--41, 2022.

\bibitem{survey2}
K.~Han, Y.~Wang, H.~Chen, X.~Chen, J.~Guo, Z.~Liu, Y.~Tang, A.~Xiao, C.~Xu, Y.~Xu \emph{et~al.}, ``A survey on vision transformer,'' \emph{IEEE Transactions on Pattern Analysis and Machine Intelligence}, pp. 87--110, 2022.

\bibitem{swinir}
J.~Liang, J.~Cao, G.~Sun, K.~Zhang, L.~Van~Gool, and R.~Timofte, ``Swinir: Image restoration using swin transformer,'' in \emph{IEEE International Conference on Computer Vision}, 2021, pp. 1833--1844.

\bibitem{sparse1}
G.~M. Correia, V.~Niculae, and A.~F. Martins, ``Adaptively sparse transformers,'' \emph{arXiv preprint arXiv:1909.00015}, 2019.

\bibitem{sparse4}
H.~Ren, H.~Dai, Z.~Dai, M.~Yang, J.~Leskovec, D.~Schuurmans, and B.~Dai, ``Combiner: Full attention transformer with sparse computation cost,'' \emph{Advances in Neural Information Processing Systems}, pp. 22\,470--22\,482, 2021.

\bibitem{sparse5}
Y.~Tay, D.~Bahri, L.~Yang, D.~Metzler, and D.-C. Juan, ``Sparse sinkhorn attention,'' in \emph{International Conference on Machine Learning}.\hskip 1em plus 0.5em minus 0.4em\relax PMLR, 2020, pp. 9438--9447.

\bibitem{sparse6}
R.~Child, S.~Gray, A.~Radford, and I.~Sutskever, ``Generating long sequences with sparse transformers,'' \emph{arXiv preprint arXiv:1904.10509}, 2019.

\bibitem{sparse7}
S.~Liu, J.~Ye, S.~Ren, and X.~Wang, ``Dynast: Dynamic sparse transformer for exemplar-guided image generation,'' in \emph{European Conference on Computer Vision}, 2022, pp. 72--90.

\bibitem{sparse10}
L.~Fan, Z.~Pang, T.~Zhang, Y.-X. Wang, H.~Zhao, F.~Wang, N.~Wang, and Z.~Zhang, ``Embracing single stride 3d object detector with sparse transformer,'' in \emph{IEEE Conference on Computer Vision and Pattern Recognition}, 2022, pp. 8458--8468.

\bibitem{sparse11}
J.~Zhang, Y.~Zhang, J.~Gu, Y.~Zhang, L.~Kong, and X.~Yuan, ``Accurate image restoration with attention retractable transformer,'' \emph{arXiv preprint arXiv:2210.01427}, 2022.

\bibitem{sparse12}
X.~Chen, H.~Li, M.~Li, and J.~Pan, ``Learning a sparse transformer network for effective image deraining,'' in \emph{IEEE Conference on Computer Vision and Pattern Recognition}, 2023, pp. 5896--5905.

\bibitem{sparse0}
S.~Chen, T.~Ye, J.~Bai, E.~Chen, J.~Shi, and L.~Zhu, ``Sparse sampling transformer with uncertainty-driven ranking for unified removal of raindrops and rain streaks,'' in \emph{IEEE International Conference on Computer Vision}, 2023, pp. 13\,106--13\,117.

\bibitem{sg1}
J.~Pan, Z.~Hu, Z.~Su, H.-Y. Lee, and M.-H. Yang, ``Soft-segmentation guided object motion deblurring,'' in \emph{IEEE Conference on Computer Vision and Pattern Recognition}, 2016, pp. 459--468.

\bibitem{sg2}
X.~Wang, K.~Yu, C.~Dong, and C.~C. Loy, ``Recovering realistic texture in image super-resolution by deep spatial feature transform,'' in \emph{IEEE Conference on Computer Vision and Pattern Recognition}, 2018, pp. 606--615.

\bibitem{sg3}
A.~Aakerberg, A.~S. Johansen, K.~Nasrollahi, and T.~B. Moeslund, ``Semantic segmentation guided real-world super-resolution,'' in \emph{Proceedings of the IEEE/CVF Winter Conference on Applications of Computer Vision}, 2022, pp. 449--458.

\bibitem{sa}
A.~Kirillov, E.~Mintun, N.~Ravi, H.~Mao, C.~Rolland, L.~Gustafson, T.~Xiao, S.~Whitehead, A.~C. Berg, W.-Y. Lo \emph{et~al.}, ``Segment anything,'' \emph{arXiv preprint arXiv:2304.02643}, 2023.

\bibitem{lin2017focal}
T.-Y. Lin, P.~Goyal, R.~Girshick, K.~He, and P.~Doll{\'a}r, ``Focal loss for dense object detection,'' in \emph{IEEE International Conference on Computer Vision}, 2017, pp. 2980--2988.

\bibitem{diceloss}
F.~Milletari, N.~Navab, and S.-A. Ahmadi, ``V-net: Fully convolutional neural networks for volumetric medical image segmentation,'' in \emph{2016 fourth international conference on 3D vision (3DV)}, 2016, pp. 565--571.

\bibitem{topk}
G.~Zhao, J.~Lin, Z.~Zhang, X.~Ren, Q.~Su, and X.~Sun, ``Explicit sparse transformer: Concentrated attention through explicit selection,'' \emph{arXiv preprint arXiv:1912.11637}, 2019.

\bibitem{topk2}
H.~Wang, J.~Shen, Y.~Liu, Y.~Gao, and E.~Gavves, ``Nformer: Robust person re-identification with neighbor transformer,'' in \emph{IEEE Conference on Computer Vision and Pattern Recognition}, 2022, pp. 7297--7307.

\bibitem{topk3}
P.~Wang, X.~Wang, F.~Wang, M.~Lin, S.~Chang, H.~Li, and R.~Jin, ``Kvt: k-nn attention for boosting vision transformers,'' in \emph{European Conference on Computer Vision}.\hskip 1em plus 0.5em minus 0.4em\relax Springer, 2022, pp. 285--302.

\bibitem{sparse13}
J.~Lin, Y.~Cai, X.~Hu, H.~Wang, Y.~Yan, X.~Zou, H.~Ding, Y.~Zhang, R.~Timofte, and L.~Van~Gool, ``Flow-guided sparse transformer for video deblurring,'' \emph{arXiv preprint arXiv:2201.01893}, 2022.

\bibitem{perceptualloss}
J.~Johnson, A.~Alahi, and L.~Fei-Fei, ``Perceptual losses for real-time style transfer and super-resolution,'' in \emph{European Conference on Computer Vision}, 2016, pp. 694--711.

\bibitem{psnrloss}
L.~Chen, X.~Lu, J.~Zhang, X.~Chu, and C.~Chen, ``Hinet: Half instance normalization network for image restoration,'' in \emph{IEEE Conference on Computer Vision and Pattern Recognition}, 2021, pp. 182--192.

\bibitem{ssimloss}
A.~Hore and D.~Ziou, ``Image quality metrics: Psnr vs. ssim,'' in \emph{IEEE International Conference on Pattern Recognition}.\hskip 1em plus 0.5em minus 0.4em\relax IEEE, 2010, pp. 2366--2369.

\bibitem{adam}
D.~P. Kingma and J.~Ba, ``Adam: A method for stochastic optimization,'' \emph{arXiv preprint arXiv:1412.6980}, 2014.

\bibitem{smith2017cyclical}
L.~N. Smith, ``Cyclical learning rates for training neural networks,'' in \emph{Proceedings of the IEEE/CVF Winter Conference on Applications of Computer Vision}.\hskip 1em plus 0.5em minus 0.4em\relax IEEE, 2017, pp. 464--472.

\bibitem{ssim}
Z.~Wang, A.~C. Bovik, H.~R. Sheikh, and E.~P. Simoncelli, ``Image quality assessment: from error visibility to structural similarity,'' \emph{IEEE Transactions on Image Processing}, vol.~13, no.~4, pp. 600--612, 2004.

\bibitem{lpips}
R.~Zhang, P.~Isola, A.~A. Efros, E.~Shechtman, and O.~Wang, ``The unreasonable effectiveness of deep features as a perceptual metric,'' in \emph{IEEE Conference on Computer Vision and Pattern Recognition}, 2018, pp. 586--595.

\bibitem{dists}
K.~Ding, K.~Ma, S.~Wang, and E.~P. Simoncelli, ``Image quality assessment: Unifying structure and texture similarity,'' \emph{IEEE Transactions on Pattern Analysis and Machine Intelligence}, vol.~44, no.~5, pp. 2567--2581, 2020.

\end{thebibliography}

\end{document}